\documentclass[12pt]{article} 
\usepackage{graphicx}
\usepackage{fix-cm}
\usepackage{mathptmx}
\usepackage{epsf}
\usepackage{amssymb}
\usepackage{amsmath}
\usepackage[dvipdfm]{hyperref}
\usepackage{theorem}  % to have rm in Def, see et.tex 
\newtheorem{theorem}{Theorem}
\newtheorem{lemma}[theorem]{Lemma}
\newtheorem{proposition}[theorem]{Proposition}
{\theorembodyfont{\rmfamily}\newtheorem{definition}[theorem]{Definition}}
\newdimen\einr
\def\abs#1{\par\hangafter=1\hangindent=\einr\noindent
\hbox to\einr{\ignorespaces#1\hfill}\ignorespaces} 
\def\reals{{\mathbb R}} 
\def\0{{\bf 0}}
\def\1{{\bf 1}}
\def\T{^{\top}} 
\def\proof{\noindent{\em Proof.\enspace}}

\def\endproof{\hfill\strut\nobreak\hfill\tombstone\par\medbreak}
\def\tombstone{\hbox{\lower.4pt\vbox{\hrule\hbox{\vrule
  \kern7.6pt\vrule height7.6pt}\hrule}\kern.5pt}}
\def\sign{{\rm sign}} 
\def\sgn{{\rm sgn}} 
\def\pf{{\rm pf}\,\,} 
\def\rooms{{\cal R}} 
\def\match{{\cal M}} 
\def\parity{{\rm parity}} 
\def\field{\textit}   % variable fields, array names
\def\method{\textsc}  % method 
\def\keyword{\textbf} % if then else etc.
\def\path{\field{path}}
\def\vc{\field{vc}}
\def\vn{\field{visitednode}}
\def\ve{\field{visitededge}}
\def\visited{\field{visited}}
\def\rank{\field{rank}}
\def\parent{\field{parent}}
\def\head{\field{head}}
\def\tail{\field{tail}}
\def\matched{\field{matched}}
\def\origmatched{\field{origmatched}}
\def\outlist{\field{outlist}}
\def\inlist{\field{inlist}}
\def\nextout{\field{nextout}}
\def\prevout{\field{prevout}}
\def\nextin{\field{nextin}}
\def\previn{\field{previn}}
\def\partner{\field{partner}}
\def\sleepcounter{\field{sleepcounter}}
\def\sleeptime{\field{sleeptime}}
\def\fosm{\method{find\_oppositely\_signed\_matching}}
\def\find{\method{find}}
\def\makeset{\method{makeset}}
\def\unite{\method{unite}}
\def\shrink{\method{shrink}}
\def\reconnect{\method{reconnect}}
\def\expandcycle{\method{expandcycle}}
\def\checkvisited{\method{checkvisited}}
\def\IF#1{\keyword{if ~}#1\keyword{~ then}}
\def\ELSE{\keyword{else}}
\def\FOR{\keyword{for}}
\def\GOTO#1{\keyword{go to }{\bf#1}}
\def\return{\keyword{return}~}

\def\SET{\leftarrow}
\def\MARK#1{\leavevmode\hbox to 3.5em{~~~\bf#1 :\hfill}\ignorespaces}
\def\3{\leavevmode\hbox to 2em{\hfill}\ignorespaces}
\def\4{\leavevmode\hbox to 3.5em{\hfill}\ignorespaces}
\newcount{\counter}
\def\5{\advance\counter by 1\noindent \leavevmode\hbox to
    1.5em{\hfill\scriptsize\the\counter}\hbox to
    2em{\hfill}\ignorespaces}
\def\6{\advance\counter by 1\noindent \leavevmode\hbox to
    1.5em{\hfill\lower.5ex\hbox{\small$^*$}\scriptsize~\the\counter}\hbox to
    2em{\hfill}\ignorespaces}
\def\institute#1{}
\begin{document}

\title{%
Oriented Euler Complexes and Signed Perfect Matchings%
}

\author{L\'aszl\'o A. V\'egh%
\thanks
{Department of Management,
London School of Economics, London WC2A 2AE, United Kingdom.
Email: L.Vegh@lse.ac.uk}
\and
Bernhard von Stengel%
\thanks
{Department of 
Mathematics,
London School of Economics, London WC2A 2AE, United Kingdom.
Email: stengel@nash.lse.ac.uk}
}

\institute{L. A. V\'egh \at
Department of Management,
London School of Economics, London WC2A 2AE, United Kingdom.
Email: L.Vegh@lse.ac.uk
\and
B. von Stengel \at
Department of Mathematics,
London School of Economics, London WC2A 2AE, United Kingdom.
Email: stengel@nash.lse.ac.uk}

\date{April 5, 2014}

\maketitle

\begin{abstract}
\noindent
This paper presents ``oriented pivoting systems'' as an
abstract framework for complementary pivoting.
It gives a unified simple proof that the endpoints of
complementary pivoting paths have opposite sign.
A special case are the Nash equilibria of a bimatrix game at
the ends of Lemke--Howson paths, which have opposite index.
For Euler complexes or ``oiks'', an orientation is defined
which extends the known concept of oriented abstract
simplicial manifolds. 
Ordered ``room partitions'' for a family of oriented oiks
come in pairs of opposite sign.
For an oriented oik of even dimension, this sign property
holds also for unordered room partitions.
In the case of a two-dimensional oik, these are perfect
matchings of an Euler graph, with the sign as defined for
Pfaffian orientations of graphs.
A near-linear time algorithm is given for the following
problem: given a graph with an Eulerian orientation with a
perfect matching, find another perfect matching of opposite
sign.
In contrast, the complementary pivoting algorithm for this
problem may be exponential.

\strut

\noindent 
\textbf{Keywords:} 
Complementary pivoting,
Euler complex,
linear complementarity problem,
Nash equilibrium,
perfect matching,
Pfaffian orientation,
PPAD.

\strut

\noindent 
\textbf{AMS 2010 subject classification:} 
90C33  %     Complementarity and equilibrium problems and variational inequalities (finite dimensions)
% 91A05     %  2-person games
% 
\end{abstract}
\vfill
\hrule
\vskip2ex
\noindent 
Apart from Appendix A and Appendix B, this article is published in 
\textit{Mathematical Programming, Series B},
DOI {10.1007/s10107-014-0770-4}, accessible online at
\url{http://link.springer.com/article/10.1007/s10107-014-0770-4}.
\newpage
\section{Introduction}

A fundamental problem in game theory is that of finding a
Nash equilibrium of a bimatrix game, that is, a two-player
game in strategic form. 
This is achieved by the classical pivoting algorithm by
Lemke and Howson (1964).
Shapley (1974) introduced the concept of an {\em index} of a
Nash equilibrium, and showed that the endpoints of every
path computed by the Lemke--Howson algorithm have opposite
index.
As a consequence, any nondegenerate game has an equal
number of equilibria of positive and negative index, if one
includes an ``artificial equilibrium'' (of, by convention,
negative index) that is not a Nash equilibrium.
The Lemke--Howson algorithm is one motivating example for
the complexity class PPAD defined by Papadimitriou (1994).
PPAD stands for ``polynomial parity argument with
direction'' and describes a class of computational problems
whose solutions are the endpoints of implicitly defined, and
possibly exponentially long, directed paths.
A salient result by Chen and Deng (2006) is that finding one
Nash equilibrium of a bimatrix game is PPAD-complete.

Lemke (1965) generalized the Lemke--Howson algorithm to
more general {\em linear complementary problems} (LCPs).
Lemke's algorithm is the fundamental {\em complementary
pivoting} algorithm;
a substantial body of subsequent work is concerned with
its applicability to LCPs and related problems (for a
comprehensive account see Cottle, Pang, and Stone, 1992).
Todd (1972; 1974) introduced a theory of 
``abstract'' complementary pivoting where the sets of basic and
nonbasic variables in a linear system are replaced by
elements of a ``primoid'' and ``duoid''.

Todd's ``semi-duoids'' have been studied independently by
Edmonds (2009) under the name of {\em Euler complexes} or
``oiks''.
A $d$-dimensional Euler complex over a finite set of {\em
nodes} is a multiset of $d$-element sets called {\em rooms}
so that any set of $d-1$ nodes is contained in an even
number of rooms.
For a family of oiks over the same node set~$V$, Edmonds
(2009) showed that there is an even number of {\em room
partitions} of~$V$, using an ``exchange algorithm'' which is
a type of parity argument.
A special case is a family of two oiks of possibly
different dimension corresponding to the two players in a
bimatrix game.
Then room partitions are equilibria, and the Lemke--Howson
algorithm is a special case of the exchange algorithm.
In another special case, all oiks in the family are the same
$2$-oik, which is an {\em Euler graph} with edges as rooms
and {\em perfect matchings} as room partitions.

This paper presents three main contributions in this context.
First, we define an abstract framework called {\em pivoting
systems} that describes ``complementary pivoting with
direction'' in a canonical manner.
Similar abstract pivoting systems have been proposed by Todd
(1976) and Lemke and Grotzinger (1976); we compare these
with our approach in Section~\ref{s-related}.
Second, using this framework, we extend the concept of
{\em orientation} to oiks and show that room partitions at
the two ends of a pivoting path have opposite sign, provided
the underlying oik is oriented.
For two-dimensional oiks, which are Euler graphs, room
partitions are perfect matchings.
Their orientation is the sign of a perfect matching as
defined for Pfaffian orientations of graphs.
Our third result is a polynomial-time algorithm for
the following problem:
Given a graph with an Eulerian orientation and a perfect
matching, find another perfect matching of opposite sign.
The complementary pivoting algorithm that achieves this may
take exponential time.

In order to motivate our general framework, we sketch here
two canonical examples (with further details in
Section~\ref{s-labpoly}) where paths of complementary pivoting
have a direction and endpoints of opposite sign.
The first example is a
simple polytope in dimension $m$ with $n$ facets, each of
which has a {\em label} in $\{1,\ldots,m\}$.
A vertex is called {\em completely labeled} if the $m$
facets it lies on together have all labels $1,\ldots,m$.
The {\em sign} of a completely labeled vertex is the sign of
the {\em determinant} of the matrix of the normal vectors of
the facets it lies on when written down in the order of
their labels.
The ``parity theorem'' states that the polytope has an equal
number of completely labeled vertices of positive and of
negative sign (so their total number is even).

The second example is that of an {\em Euler digraph} with
vertices $1,\ldots,m$ and edges oriented so that each node
of the graph has an equal number of incoming and outgoing
edges.
A perfect matching of this graph has a sign obtained as
follows: Consider any ordering of the matched edges and
write down the two endpoints of each matched edge in the
order of its orientation. 
This defines a permutation of the nodes, whose {\em parity}
(even or odd number of inversions) defines the sign of the
matching.  
Here the ``parity theorem'' states that the Euler digraph
has an equal number of perfect matchings of positive and of
negative sign.

The first example is a case of a ``vertical''  LCP (Cottle
and Dantzig, 1970) and the second of an oik partition.
Both parity theorems have a canonical proof where the
completely labeled vertices and perfect matchings,
respectively, are connected by paths of ``almost
completely labeled vertices'' or ``almost matchings'',
respectively.
The orientation of the path uses that exchanging two 
columns of a determinant switches its sign, and that
exchanging two positions in a permutation switches its
parity.
In addition, one has to consider how the ``pivoting''
operation changes such signs.
Our concept of a pivoting system (see
Definition~\ref{d-pivot}) takes account of these features
while keeping the canonical proof.

In Section~\ref{s-labpoly} we describe our two motivating
examples in more detail.
Labeled polytopes and their completely labeled (``CL'')
vertices are related to LCPs, and
are equivalent to equilibria in bimatrix games
(Proposition~\ref{p-unitv}).
We also give a small example of the pivoting algorithm that
finds a second perfect matching in an Euler digraph.

In Section~\ref{s-compl} we describe our framework of
oriented pivoting systems, and prove the main ``parity''
Theorem~\ref{t-sign}.
The section concludes with the application to labeled
polytopes.

We study orientation for oiks in Section~\ref{s-oiks}.
The general Definition~\ref{d-oikorient} seems to be a new
concept, which extends the known orientation for abstract
simplicial manifolds (e.g., Hilton and Wylie, 1967; Lemke
and Grotzinger, 1976) and ``proper duoids'' (Todd, 1976).
Then the parity theorem applies to ordered room partitions in
oriented oiks, where the order of rooms in a partition is
irrelevant for oiks of even dimension; see
Theorem~\ref{t-ordpart} and Theorem~\ref{t-part}.

Section~\ref{s-related} discusses related work, in
particular of Todd (1972; 1974; 1976) and of Edmonds (2009)
and Edmonds, Gaubert, and Gurvich (2010).

Section~\ref{s-matchings} is concerned with signed perfect
matchings in Euler digraphs.
A second perfect matching of opposite sign is guaranteed to
exist by the complementary pivoting algorithm, which,
however, may take exponential time.
In Theorem~\ref{t-euler} we give an algorithm to find such
an oppositely signed matching in near-linear time in the
number of edges of the graph.
This is closely related to the well-studied theory of
Pfaffian orientations: an orientation of an undirected graph
is Pfaffian if all perfect matchings have the same sign.
It is easy to see directly that an Euler digraph is not
Pfaffian; our result can be seen as a constructive
and computationally efficient verification of this fact.

Issues of computational complexity are discussed in the
concluding Section~\ref{s-conclude}.

\section{Labeled polytopes and signed matchings}
\label{s-labpoly}

In this preliminary section, we present two main examples
that we generalize later in an abstract framework.
The first example is a labeled polytope, whose completely
labeled (``CL'') vertices provide an intuitive geometric
view of Nash equilibria in a bimatrix game.
We also mention the connection to the linear complementarity
problem.
The second example is an Euler digraph with its perfect
matchings.

We use the following notation.
Let $[k]=\{1,\ldots,k\}$ for any positive integer $k$.
The transpose of a matrix $B$ is $B^\top$.
All vectors are column vectors.
The zero vector is~$\0$, the vector of all ones is $\1$,
their dimension depending on the context. 
Inequalities like $x\ge\0$ between two vectors hold for all
components.
A {\em unit vector} $e_k$ has its $k$th component equal to
one and all other components equal to zero.
A permutation $\pi$ of $[m]$ has {\em parity} $(-1)^k$ if
$k$ is the number of its {\em inversions}, that is, pairs
$i,j$ so that $i<j$ and $\pi(i)>\pi(j)$, and the permutation
is also called even or odd when $k$ is even or odd,
respectively.

A polyhedron $P$ is the intersection of $n$ halfspaces in
$\reals^m$,
\begin{equation}
\label{P}
P=\{x\in\reals^m\mid a_j\T x\le b_j,~j\in[n]\}
\end{equation}
with vectors $a_j$ in $\reals^m$ and reals $b_j$.
A {\em labeling function} $l:[n]\to[m]$ assigns a label to
each inequality in (\ref{P}), and $x$ in $P$ is said to have
label $l(j)$ when the $j$th inequality is binding, that is,
$a_j\T x=b_j$, for any $j$ in~$[n]$.
The polyhedron $P$ is a polytope if it is bounded.
A {\em vertex} of $P$ is an extreme point of $P$, that is, a
point that cannot be represented as a convex combination of
other elements of~$P$.

We normally look at ``nondegenerate'' polytopes where
binding inequalities define facets, and no more than $m$
inequalities are ever binding.
That is, we assume $P$ is a simple polytope 
(every vertex lies on exactly $m$ facets) and that
none of the inequalities can be omitted without changing
the polytope, so for every $j$ in $[n]$ the $j$th binding
inequality defines a facet $F_j$ given by 
\begin{equation}
\label{Fj}
F_j= \{x\in P\mid a_j\T x=b_j\}
\end{equation}
(for notions on polytopes see Ziegler, 1995).
Then facet $F_j$ has label $l(j)$ for $j$ in $[n]$,
and we call $P$ a {\em labeled polytope}.
A vertex of $P$ is {\em completely labeled} or {\em CL} if
the $m$ facets it lies on have together all labels in $[m]$.

CL vertices of polytopes are closely related to Nash
equilibria in bimatrix games.
Suppose the polytope $P$ has the form
\begin{equation}
\label{Pge0}
P=\{x\in\reals^m\mid -x\le\0,~C x\le \1\}
\end{equation}
for some $(n-m)\times m$ matrix $C$, and that each of the
first $m$ inequalities $x_i\ge0$ has label $i$ in~$[m]$.
Then $\0$ is a completely labeled vertex.
If $P$ in (\ref{P}) has a completely labeled vertex, then it
is easy to see that it can be brought into the form
(\ref{Pge0}) by a suitable affine transformation that maps
that vertex to~$\0$ (see von Stengel, 1999, Prop.~2.1).
If $C$ is a square matrix, then the CL vertices $x$ of $P$
other than $\0$ correspond to symmetric Nash equilibria
$(\hat x,\hat x)$ of the symmetric game with payoff matrices
$(C,C\T)$, where $\hat x=x/\1\T x$.
In turn, symmetric equilibria of symmetric games encode Nash
equilibria of arbitrary bimatrix games (see, e.g., Savani
and von Stengel, 2006, also for a description of the
Lemke--Howson method in this context).
Hence, given a bimatrix game, its Nash equilibria are
encoded by the CL vertices (other than $\0$) of a polytope
$P$ in~(\ref{Pge0}).

Conversely, consider a labeled polytope $P$ with a CL
vertex $\0$ as in~(\ref{Pge0}).
For a general matrix $C$ in (\ref{Pge0}) and general labels
for the inequalities $Cx\le\1$, the following proposition
implies that the CL vertices of $P$ correspond to Nash
equilibria of a ``unit-vector game'' $(A,C\T)$.
The unit vectors that form the columns of $A$ encode the
labels for the inequalities $Cx\le\1$.
(This proposition holds even if a point of $P$ may have more
than $m$ binding inequalities, except that then a CL point
of~$P$ is not necessarily a vertex.)
The proposition, in a dual version, was first stated and
used by Balthasar (2009, Lemma 4.10).
The special case when $A$ is the identity matrix describes 
an ``imitation game'' whose equilibria correspond to the
symmetric equilibria of the symmetric game $(C,C\T)$
(McLennan and Tourky, 2010).
For further connections see Section~\ref{s-related}.

\begin{proposition}
\label{p-unitv}
Suppose that $(\ref{Pge0})$ defines a polytope $P$
so that the inequalities $-x_i\le0$ have label $i$ for 
$i\in[m]$, and the last $n-m$ inequalities $Cx\le\1$ have
labels $l(m+j)$ in $[m]$ for $j\in[n-m]$.
Then $x$ is a CL point of $P-\{\0\}$ if and only if
for some~$\hat y$ the pair $(x/(\1\T x),\hat y)$ 
is a Nash equilibrium of the $m\times (n-m)$ game $(A,C\T)$
where $A=[e_{l(m+1)}\cdots e_{l(n)}]$.
\end{proposition}

\proof
Consider the game $(A,C\T)$ as described.
Then a mixed strategy pair $(\hat x,\hat y)$ with payoffs $u$ to player~1
and $v$ to player~2 is a Nash equilibrium if and only if
\begin{equation}
\label{AyCx}
\hat x\ge\0,
\quad
\1\T \hat x=1,
\quad
\hat y\ge\0,
\quad
\1\T \hat y=1,
\quad
A\hat y\le\1u,
\quad
C\hat x\le\1v, 
\end{equation}
and the ``best response'' (or complementarity) conditions
\begin{equation}
\label{comp}
\forall i\in[m]~:~ \hat x_i>0 \Rightarrow (A\hat y)_i=u,
\qquad
\forall j\in[n]~:~ \hat y_j>0 \Rightarrow (C\hat x)_j=v
\end{equation}
hold.
Condition (\ref{AyCx}) implies $u>0$ and $v>0$, as follows.
First, $\hat y_j>0$ for some~$j$ in $[n]$.
The $j$th column of $A$ is the unit vector $e_i$ for
$i=l(m+j)$, so for the $i$th row $(A\hat y)_i$ of $A\hat y$
we have $u\ge(A\hat y)_i\ge \hat y_j>0$.
Second, if $v\le0$ then $C\hat x\le\1v\le\0\le\1$, and hence
$C\hat x\lambda\le\1$ for any real $\lambda\ge0$, where $\hat x\ne\0$,
so that $P$ contains the infinite ray
$\{\hat x\lambda\mid\lambda\ge0\}$, but $P$ is bounded.
So indeed $u>0$ and $v>0$.
With $x=\hat x/u$ and $y=\hat y/v$, conditions
(\ref{AyCx}) and (\ref{comp}) are equivalent to 
\begin{equation}
\label{AyCx1}
x\ge\0,
\quad
x\ne\0,
\quad 
y\ge\0,
\quad
y\ne\0,
\quad 
Ay\le\1,
\quad
Cx\le\1, 
\end{equation}
and
\begin{equation}
\label{comp1}
\forall i\in[m]~:~ x_i>0 \Rightarrow (Ay)_i=1,
\qquad
\forall j\in[n]~:~ y_i>0 \Rightarrow (Cx)_j=1,
\end{equation}
from which (\ref{AyCx}) and (\ref{comp}) are obtained with
$u=1/\1\T x$,
$v=1/\1\T y$,
$\hat x=xu$,
$\hat y=yv$.  

Suppose now that $(\hat x,\hat y)$ is an equilibrium, with
$x$ and $y$ so that (\ref{AyCx1}) and (\ref{comp1}) hold.
Then $x\in P$ and we want to show that $x$ is a CL point
of~$P$.
Let $i\in[m]$.
If $x_i=0$ then $x$ has label~$i$, so let $x_i>0$. 
Then $(Ay)_i=1$ by (\ref{comp1}), so there is some $j$
in~$[n]$ so that the $j$th column of $A$ is $e_i$, that is,
$l(m+j)=i$, and $y_j>0$.
By (\ref{comp1}), $(Cx)_j=1$, so the $j$th inequality in
$Cx\le\1$ is binding, which has label $l(m+j)=i$.
So $x$ is CL.

Conversely, let $x$ be a CL point of~$P$ and $x\ne\0$.
Then for each $i$ in $[m]$ with $x_i>0$, label $i$ for $x$
comes from a binding inequality $(Cx)_j=1$ with label
$l(m+j)=i$, so we let $y_j=1$ for the smallest $j$ with this
property, and set $y_k=0$ for all other $k$ in~$[n]$.
Then $x_i>0$ implies $(Ay)_i=(e_i)_i=1$, and $y_j>0$ implies
$(Cx)_j=1$, so (\ref{AyCx1}) and (\ref{comp1}) hold,
and with $\hat x=x/\1\T x$ and $\hat y=y/\1\T y$ we obtain
the Nash equilibrium $(\hat x,\hat y)$ of $(A,C\T)$.
\endproof

A {\em linear complementarity problem} (LCP) with
an $m\times m$ matrix $M$ and $m$-vector $q$ is the problem
of finding $z$ in $\reals^m$ so that $z\ge\0$, $q+Mz\ge\0$,
and $z\T(q+Mz)=0$ (see Cottle, Pang, and Stone, 1992).
This the same as finding a CL point $z$ of the polyhedron
\begin{equation}
\label{lcp}
P=\{z\in\reals^m\mid -z\le\0,~ -Mz\le q\,\}
\end{equation}
whose $2m$ inequalities have labels $1,\ldots,m,1,\ldots,m$.
More generally, $M$ in (\ref{lcp}) may be of size $(n-m)\times m$
with labels $1,\ldots,m$ for the inequalities $-z\le\0$ and
arbitrary labels in $[m]$ for the inequalities $-Mz\le q$. 
This is known as the ``vertical'' LCP (Cottle and Dantzig,
1970).
Lemke (1965) described a path-following method of
``complementary pivoting'' to solve LCPs; many studies
concern whether this method terminates depending on $M$ and~$q$.
This is always so in our special case (\ref{Pge0}) where
$q=\1$, $C=-M$, and $P$ is bounded.

For a simple polytope $P$ in (\ref{Pge0}), a ``Lemke path''
(see also Morris, 1994) is obtained for a given {\em missing label}~$w$
in $[m]$ as follows.
Start at a CL vertex, for example $\0$, and ``pivot'' along
the unique edge that leaves the facet with label~$w$.
This reaches a vertex $v$ on a new facet $F$ with label~$k$.
If $k=w$, then $v$ is CL and the path terminates.
Otherwise, label $k$ is {\em duplicate}, that is, $v$ is on
another facet that also has label $k$.
Continue by pivoting away from that facet to the next
vertex, which again has a label that is either $w$ or
duplicate, and repeat.
This defines a unique path that consists of vertices and
edges all of which have all labels except~$w$, and whose
endpoints are CL.
The CL vertices of $P$ are the unique endpoints of these
``Lemke paths'' and hence there is an even number of them,
which is the basic ``parity theorem''.
In addition, a CL vertex has a {\em sign} which is the sign
of the determinant of the $m$ normal vectors of the binding
inequalities when these vectors are written down in the
order of their labels $1,\ldots,m$.
Then the endpoints of a Lemke path have opposite sign,
as essentially shown by Shapley (1974).
We prove this in more general form in Theorem~\ref{t-sign}
and Proposition~\ref{p-det} below.

Our second example is given by an Euler digraph $G=(V,E)$,
that is, a graph so that each edge is
oriented so that every node of $G$ has equally many incoming
and outgoing edges.
We allow multiple parallel edges between two nodes.
Let $V=[m]$.
A {\em perfect matching} $M$ is a set of $m/2$ edges no two
of which have a node in common.
The {\em sign} of a perfect matching $M$ is defined as
follows.
Consider the edges in $M$ in some order and write down
their endpoints in the order of the orientation of the edge.
This defines a permutation of $V$.
The sign of $M$ is the parity of that permutation, which
is independent of the order of edges.

\begin{figure}[hbt] 
\strut\hfill
\epsfxsize3.3cm
\epsfbox{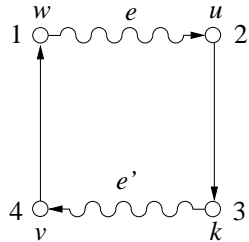} 
\hfill
\strut 
\caption{Example of an Euler digraph and
matched edges (wiggly lines) $(1,2)$ and $(3,4)$.
Here all edges are uniquely paired because every node has only
one incoming and one outgoing edge.
} 
\label{Gpivot}
\end{figure}

A ``pivoting path'' that starts from a perfect matching $M$
of $G$, and finds a second perfect matching, can be defined
as follows (see Figure~\ref{Gpivot} for a simple example).
Choose a {\em missing} node~$w$ and for each node of $G$ a
fixed {\em pairing} between its incoming and outgoing edges.
Let $e$ be the matched edge incident to $w$, for example
oriented from $w$ to $u$, so $e=(w,u)$.
Consider the (necessarily unmatched) edge $(u,k)$ at the other
endpoint $u$ of $e$ that is paired with~$e$.
(If $e$ was oriented as $(u,w)$, the paired edge would be
$(k,u)$.) 
Replace $e$ in $M$ with $(u,k)$.
Unless $k=w$, the result is an ``almost matching'' with a
node $k$ in $V$ that is incident to two edges $(u,k)$ and
$e'$, say, node $w$ that is not incident to any edge, and
every other node incident to exactly one edge.
Consider the endpoint $v$ of $e'$ other than $k$, and
(assuming $e'$ is oriented as $e'=(k,v)$), replace $e'$ again
with its paired unmatched edge $(v,x)$ at $v$ (in
Figure~\ref{Gpivot}, $x=w$).
Continue in this manner until the endpoint of the newly
found edge is~$w$.
It can be shown that the matching of $G$ that is found has
opposite sign to the original matching.
In Figure~\ref{Gpivot}, the two matchings are $\{12,34\}$ and 
$\{23,41\}$ which have indeed opposite sign.

\section{Labeled oriented pivoting systems}
\label{s-compl}

In this section we describe a general abstract framework of
``complementary pivoting'' with orientation.
We will use an abstract set of {\em states} (which may be
vertices of a polytope, or sets of edges, such as matchings,
in a digraph) and their {\em representations} which define how
to assign labels, orientations, signs, and how to {\em
pivot} from one state to another.

Consider a finite set $S$ of {\em states}.
Each state $s$ is {\em represented} by an $m$-tuple 
\begin{equation}
\label{rs}
r(s)=(s_1,\ldots,s_m)
\end{equation}
of {\em nodes} $s_i$ from a given set $V$.
For a polytope as in (\ref{P}), the set of nodes $V$ is the
set $[n]$ that numbers its facets, and a state is a vertex
of~$P$ represented by the $m$ facets it lies on.
In an Euler digraph, $V$ is the set of $m$ nodes of
the graph, and a state $s$ is a set of $m/2$ edges.
A representation of $s$ is an $m$-tuple $(s_1,\ldots,s_m)$
so that the oriented edges in $s$ are $(s_{2i-1},s_{2i})$ 
for $1\le i\le m/2$.
Note that this representation may not identify $s$ uniquely
if the graph has parallel edges.  

The pivoting operation $f$ takes a state $s$ and $i$ in
$[m]$ and produces a new state $t$, with the effect that
the $i$th component $s_i$ of the representation of $s$ in
(\ref{rs}) is replaced by another element $u$ of $V$.
We denote the resulting $m$-tuple by $(r(s)\mid i\to u)$, 
\begin{equation}
\label{tou}
((s_1,\ldots,s_m)\mid i \to u)
= (s_1,\ldots,s_{i-1},u,s_{i+1},\ldots,s_m).
\end{equation}
We denote the resulting new state with this representation
by $t=f(s,i)$.
The pivoting step is simply reversed by $s=f(t,i)$.
(We will soon refine this by allowing $r(t)$ to be a
permutation of $r(s)$.)
In the polytope, $s$ and $t$ are adjacent vertices that
agree in all binding inequalities except for the $i$th one.

In an Euler digraph with paired incoming and outgoing edges
at each node, an example of pivoting is the following:
Suppose state $s$ (set with $m/2$ edges) has the edge
$(s_1,s_2)$, which is paired with $(s_2,u)$ in the graph, and let
$i=1$.
Pivoting replaces $(s_1,s_2)$ with $(s_2,u)$, giving the new
state~$t$.
Here we encounter the difficulty that the representations of
$s$ and $t$ should be $r(s)=(s_1,s_2,s_3,\ldots,s_m)$ and
$r(t)=(s_2,u,s_3,\ldots,s_m)$ in order to write down the
edges in their orientation. 
However, this requires that $s_2$ appears in a permuted
place from $r(s)$ in $r(t)$; this is addressed in 
Definition~\ref{d-pivot} below which is more general than
the description so far.

Each node $u$ in $V$ has a {\em label} $l(u)$ given by a
labeling function $l:V\to[m]$.
The path-following argument has as endpoints of the paths
{\em completely labeled (CL)} states $s$ where, given
(\ref{rs}), $\{l(s_i)\mid i\in[m]\}=[m]$.
In addition, it considers states $s$ that are {\em
almost completely labeled (ACL)} defined by the condition
$\{l(s_i)\mid i\in[m]\}=[m]-\{w\}$, where $w$ is called the
{\em missing label} and the unique $k$ so that
$k=l(s_i)=l(s_j)$ for $i\ne j$ is called the {\em duplicate
label}.  

``Complementary pivoting'' means the following:
Start from a CL state $s$ and allow a specific label $w$ to
be missing, where $l(s_i)=w$.
Pivot to the state $t=f(s,i)$.
Then if the new node $u$ in (\ref{tou}) has label $l(u)=w$,
then $t$ is CL and the path ends.
Otherwise, $l(u)$ is duplicate, with $l(s_j)=l(u)$ for
$j\ne i$, so that the next state is obtained by pivoting 
to $f(t,j)$, and the process is repeated.
This defines a unique path that starts with a CL state,
follows a sequence of ACL states, all of which have missing
label~$w$, and ends with another CL state.
The path cannot meet itself because the pivoting function is
invertible; hence, the process terminates.

We also want to give a {\em direction} to the pivoting path.
For this purpose, a CL state will get a {\em sign}, either
$+1$ or $-1$, so that the two CL states at the ends of the
path have opposite sign.
This sign is the product of two such numbers (again either
$+1$ or $-1$), namely the {\em orientation} $\sigma(s)$ of
the state $s$ when represented as $r(s)=(s_1,\ldots,s_m)$,
and the parity of the permutation $\pi$ of $[m]$ when
writing down the nodes $s_1,\ldots,s_m$ in ascending order
of their labels.  
In the polytope setting, the orientation of a vertex is the
sign of the determinant of the normal vectors $a_j$ of the
facets $F_j$ that contain that vertex, see (\ref{det}) below.
The important abstract property is that pivoting from
$(s_1,\ldots,s_m)$ to $((s_1,\ldots,s_m)\mid s_i\to u)$
changes the orientation, stated for polytopes in
Proposition~\ref{p-det} below.

In order to motivate the following definition, we first give
a very simple example of a pivoting path with only one ACL
state apart from its two CL states at its ends.
Consider $V=\{a_1,a_2,a_3,b_1,b_2\}$ with labels
$l(a_1)=l(b_1)=1$, $l(a_2)=l(b_2)=2$, $l(a_3)=3$,
and three states $s^0,s^1,s^2$ with
$r(s^0)=(a_1,a_2,a_3)$,
$r(s^1)=(b_2,a_2,a_3)$,
$r(s^2)=(b_2,b_1,a_3)$.
Assume that $f(s^0,1)=s^1$ and $f(s^1,2)=s^2$.
Then starting from the CL state $s^0$ and missing label $1$
pivots to $s^1$ (by replacing $a_1$ with $b_2$), which is an
ACL state with duplicate label~$2$ in the two positions $1$
and $2$.
The next complementary pivoting step pivots from $s^1$ to
$s^2$ (by replacing $a_2$ with $b_1$), where $s^2$ is CL and
the path ends.
The three states have the following orientations:
$\sigma(s^0)=1$, $\sigma(s^1)=-1$, $\sigma(s^2)=1$,
which alternate as one state is obtained from the next by
pivoting.
Here, the two CL states $s^0$ and $s^2$ have the same
orientation.
They obtain their {\em sign} by writing their nodes in
ascending order of their labels:
This is already the case for $r(s^0)$, but in $r(s^2)$ the
permutation $2,1,3$ of the labels is odd, so the sign of
$s^2$ becomes $-1$, which is indeed opposite to the sign of
$s^0$.

In this example, we have chosen the representations of the
states $s^0,s^1,s^2$ in such a way that the required
pivoting steps can indeed be performed by exchanging a node
at a fixed position; however, this may not be clear in
advance: another representation of the three states might be
$(a_1,a_2,a_3)$, $(a_2,a_3,b_2)$, $(a_3,b_1,b_2)$.
In this case, we still allow pivoting from $s^0$ to $s^1$
by going from $(a_1,a_2,a_3)$ to $(b_2,a_2,a_3)$ but with a
subsequent, known permutation $\pi$ to obtain the
representation $(a_2,a_3,b_2)$ of $s^1$;
for a ``coherent'' orientation of the states, we have to
take the parity of $\pi$ into account.

\begin{definition}
\label{d-pivot}
A {\em pivoting system} is given by $(S,V,m,r,f)$ with a finite
set $S$ of {\em states}, a finite set $V$ of {\em nodes}, a
positive integer $m$,
a {\em representation} function $r:S\to V^m$, 
and a {\em pivoting function} $f:S\times[m]\to S$. 
For a permutation $\pi$ of $[m]$ and
$r(t)=(t_1,\ldots,t_m)$,
let 
\begin{equation}
\label{rpi}
r^\pi(t)=(t_{\pi(1)},\ldots,t_{\pi(m)}).
\end{equation}
Then for each $t=f(s,i)$, there is a permutation
$\pi=\pi(s,i)$ of
$[m]$ so that
$r^\pi(t)=(r(s)\mid i\to u)$
for some $u$ in $V$, and $f(t,\pi(i))=s$.
The pivoting system is {\it oriented} if each state $s$
has an {\em orientation} $\sigma(s)$, where
$\sigma:S\to\{-1,1\}$, so that
\begin{equation}
\label{switch}
\sigma(t)=-\sigma(s)\cdot \parity(\pi)
\end{equation}
whenever $t=f(s,i)$ with $\pi=\pi(s,i)$ as described.
\end{definition}

Note that when pivoting from state $s$ to state $t=f(s,i)$,
the permutation $\pi$ so that $r^\pi(t)=(r(s)\mid i\to u)$
is a function $\pi(s,i)$ of $s$ and $i$ and hence part of
the pivoting system.
In addition, the orientation $\sigma$ of the states is
unique only up to possible multiplication with $-1$; usually
one of the two possible orientations that are ``coherent''
according to (\ref{switch}) is chosen as a convention
(for Nash equilibria of bimatrix games, for example, so that
the CL vertex $\0$ of $P$ in (\ref{Pge0}) has negative
sign).

The following simple example illustrates the use of the
permutation $\pi=\pi(s,i) $ in Definition~\ref{d-pivot}.
Suppose % $V$ contains integers and
$r(s)=(s_1,s_2,s_3)=(1,2,3)$ and
$r(t)=(t_1,t_2,t_3)=(2,3,4)$, where $f(s,1)=t$
by replacing $s_1$ with~$4$.
This means that
$r^{\pi}(t)=(t_{\pi(1)},t_{\pi(2)},t_{\pi(3)})=
((s_1,s_2,s_3)\mid 1\to4)=(4,2,3)$, so
$\pi(1)=3$, 
$\pi(2)=1$, 
$\pi(3)=2$, 
that is, $\pi$ says that $s_j$ becomes $t_{\pi(j)}$
except for the ``pivot element''~$s_i$.
Pivoting ``back'' gives $s=f(t,\pi(1))=f(t,3)$.

It is important to note that the pivot operation $f$
operates on states $s$ which gives a new state $t=f(s,i)$, where
$i$ refers to the $i$th component $s_i$ of the representation
$r(s)=(s_1,\ldots,s_m)$.
However, there may be different states $s$ and $s'$ with the
same representation $r(s)=r(s')$, as we will see in later 
examples; otherwise, we could just take $S$ as a subset
of~$V^m$ and dispense with~$r$.
This is one distinction to the formal approaches of
Lemke and Grotzinger (1976) and Todd (1976), who, in
addition, assume that the nodes $s_1,\ldots,s_m$ in
(\ref{rs}) are distinct, which we do not require either.
Furthermore, we do not give signs to the two equivalence
classes of even and odd permutations of $(s_1,\ldots,s_d)$,
as Hilton and Wylie (1967) or Todd (1976), but instead
consider unique representations $r(s)$, and build a single
permutation $\pi$ into each pivoting step.

The pivoting system $(S,V,m,r,f)$ is {\em labeled} if there 
is a labeling function $l:V\to[m]$.
For $(s_1,\ldots,s_m)$ where $s_i\in V$ for $i$ in $[m]$,
let  $l(s_1,\ldots,s_m)=(l(s_1),\ldots,l(s_m))$, and
consider this $m$-tuple as a permutation of $[m]$ if
$l(s_i)\ne l(s_j)$ whenever $i\ne j$.
If the pivoting system is oriented, then the {\em sign} of a
CL state $s$ is defined as 
\begin{equation}
\label{CLsign}
\sign(s)=\sigma(s)\cdot \parity(l(r(s))).
\end{equation}
For an ACL state $s$, we define {\em two} opposite signs as
follows: consider the positions $i,j$ of the duplicate label
in $r(s)=(s_1,\ldots,s_m)$, that is, $l(s_i)=l(s_j)$ with
$i\ne j$, and missing label~$w$.
Replacing $l(s_i)$ with $w$ in $l(r(s))$ then defines
a permutation of $[m]$, denoted by $(l(r(s))\mid i\to w)$,
which has opposite parity to $(l(r(s))\mid j\to w)$ because
that permutation is obtained by switching the labels $w$ and
$l(s_j)$ in positions $i$ and $j$.
Let
\begin{equation}
\label{ACLsign}
\sign(s,i)=\sigma(s)\cdot \parity(l(r(s))\mid i\to w),
\end{equation}
so
\begin{equation}
\label{opposite}
\sign(s,j)=\sigma(s)\cdot \parity(l(r(s))\mid j\to w)
=-\sign(s,i).
\end{equation}
This is the basic observation, together with the
orientation-switching of a pivoting step stated in (\ref{switch}),
to show that complementary pivoting paths in an oriented
pivoting system have a direction.
This direction (say from negatively to positively signed CL
end-state) is also locally recognized for any ACL state on
the path, as stated in the following theorem.
Hence, for a fixed missing label $w$, the endpoints of the
paths define pairs of CL states of opposite sign.
The pairing may depend on $w$, but the sign of each CL state
does not.

\begin{theorem}
\label{t-sign}
Let $(S,V,m,r,f)$ be a pivoting system with a
labeling function $l:V\to[m]$, and fix $w\in[m]$.
\abs{(a)}
The CL states and ACL states with missing label~$w$ are
connected by complementary pivoting steps and form a set of
paths and cycles, with the CL states as endpoints of the
paths.
The number of CL states is even.
\abs{(b)}
Suppose the system is oriented.
Then the two CL states at the end
of a path have opposite sign.
When pivoting from an ACL state $s$ on that path to
$t=f(s,i)$ where $l(s_i)$ is the duplicate label in
$r(s)=(s_1,\ldots,s_m)$, the CL state found at the end
of the path by continuing to pivot in that direction has
opposite sign to $\sign(s,i)$.  
There are as many CL states of sign $1$ as of sign~$-1$.
\end{theorem}

\proof
Assume that the pivoting system is oriented; otherwise
complementary pivoting (already described informally above) 
is part of the following description by disregarding all
references to signs.
Consider a CL state $s$ and $r(s)=(s_1,\ldots,s_m)$, with
$w$ declared as the missing label for the path that starts
at $s$, and let $l(s_i)=w$.
We can define $\sign(s,i)$ as in (\ref{ACLsign}), which is
just $\sign(s)$ in (\ref{CLsign}), because
$l(r(s))=(l(r(s))\mid i\to w)$.
The following considerations apply in the same way if $s$ is
an ACL state with duplicate label $l(s_i)$.
The path starts (or continues, if $s$ is ACL) by pivoting to
$t=f(s,i)$.
Assume $r^\pi(t)=(r(s)\mid i\to u)$ as in
Definition~\ref{d-pivot}.
Then $(l(r(s))\mid i\to w)$ is a permutation of $[m]$, which
is equal to 
$(l(r^\pi(t))\mid i\to w)$, and
$(l(r(t))\mid \pi(i)\to w)$ is a permutation of $[m]$
with $\parity(\pi)\cdot \parity(l(r(s))\mid i\to w)$ as its
parity.
Hence, by~(\ref{switch})
\[
\arraycolsep0pt
\begin{array}{rl}
 \sign(s,i)
 &{}=\sigma(s)\cdot \parity(l(r(s))\mid i\to w)\\
 &{}=-\sigma(t)\cdot \parity(\pi)\cdot \parity(l(r(s))\mid i\to w)\\
 &{}=-\sigma(t)\cdot \parity(l(r(t))\mid \pi(i)\to w)\\
 &{}=-\sign(t,\pi(i)).\\
\end{array}
\]
If $l(u)$ is the missing label $w$, then $t$ is the CL state
at the other end of the path and $\sign(t)=\sign(t,\pi(i))$,
which is indeed the opposite sign of the starting state~$s$.
Otherwise, label $l(u)$ is duplicate, with $l(u)=l(s_j)$
for some $j\ne i$, that is, $l(t_{\pi(i)})=l(t_{\pi(j)})$
for $r(t)=(t_1,\ldots,t_m)$,
so that the path continues with the next pivoting step from
$t$ to $f(t,\pi(j))$, where by (\ref{opposite}) 
\[
\sign(t,\pi(j))=-\sign(t,\pi(i))=\sign(s,i),
\]
that is, this step continues from a state with the same sign
as the starting CL state, and the argument repeats.  
This proves the theorem.
\endproof

For a labeled polytope $P$ as in (\ref{P}), an oriented
pivoting system is obtained as follows:
The states in $S$ are the vertices $x$ of $P$, and by the
assumptions on $P$ each vertex $x$ lies on exactly $m$
facets $F_{s_1},\ldots,F_{s_m}$, where we take
$r(x)=(s_1,\ldots,s_m)$ as the representation of $x$ with
$s_1,\ldots,s_m$ in any fixed (for example, increasing)
order.
Moreover, the normal vectors $a_{s_1},\ldots,a_{s_m}$ of
these facets in (\ref{Fj}) are linearly independent.
For any $i$ in $[m]$, the set
$\bigcap_ {p\in[m]-\{i\}} F_{s_p}$
is an edge of $P$ with two vertices $x$ and $y$ as its
endpoints, which defines the pivoting function as
$y=f(x,i)$.
The orientation of the vertex $x$ is given by
\begin{equation}
\label{det}
\sigma(x)=\sgn(\det[a_{s_1}\cdots a_{s_m}])
\end{equation}
with the usual sign function $\sgn(z)$ for reals $z$ and the
determinant $\det\,A$ for any square matrix $A$.
The following proposition is well known (Lemke and
Grotzinger, 1976, for example, argue with linear programming
tableau entries; Eaves and Scarf, 1976, Section~5, consider
the index of mappings); we give a short geometric proof.

\begin{proposition}
\label{p-det}
A labeled polytope $P$ with orientation $\sigma(x)$ as in
$(\ref{det})$ for each vertex $x$ of $P$ defines an oriented
pivoting system.
\end{proposition}

\proof
Consider pivoting from $x$ to vertex $y=f(x,i)$.
We want to prove (\ref{switch}), that is,
$\sigma(y)=-\sigma(x)\cdot\parity(\pi)$ where $\pi$ is the
permutation so that $r^\pi(y)=(r(x)\mid i\to u)$.
Let $x$ be on the $m$ facets $F_{s_1},F_{s_2},\ldots,F_{s_m}$ as in
(\ref{Fj}).
The representation $r(x)=(s_1,\ldots,s_m)$ determines
the order of the columns of the matrix
$[a_{s_1}a_{s_2}\cdots a_{s_m}]$ whose determinant
determines the orientation $\sigma(x)$ in (\ref{det}).
Any permutation of the columns of this matrix changes
the sign of the determinant according to the parity
of the permutation, so for proving (\ref{switch}) the actual
order of $(s_1,\ldots,s_m)$ in $r(x)$ does not matter as
long as it is fixed.
Hence, we can assume that $\pi$ is the identity permutation,
and that pivoting affects the first column ($i=1$), so that
$y$ is on the $m$ facets $F_{s_0},F_{s_2},\ldots,F_{s_m}$.

We show that $\det[a_{s_0}a_{s_2}\cdots a_{s_m}]$
and $\det[a_{s_1}a_{s_2}\cdots a_{s_m}]$ have opposite sign,
that is, $\sigma(y)=-\sigma(x)$ as claimed.
The $m+1$ vectors $a_{s_0},a_{s_1},a_{s_2},\ldots,a_{s_m}$
are linearly dependent, so there are reals
$c_0,c_1,\ldots,c_m$, not all zero, with
\begin{equation}
\label{lindep}
\sum_{p=0}^m c_p\, a_{s_p}\T=\0\T.
\end{equation}
Note that $c_0\ne0$, because otherwise the normal vectors
$a_{s_1},a_{s_2},\ldots,a_{s_m}$ of the facets that define
$x$ would be linearly dependent, and similarly $c_1\ne0$.
Multiply the sum in (\ref{lindep}) with both $y$ and $x$, 
where $a_{s_p}\T y=a_{s_p}\T x=b_{s_p}$ for $p=2,\ldots,m$.
This shows
$c_0\,a_{s_0}\T y+c_1\,a_{s_1}\T y=
c_0\,a_{s_0}\T x+c_1\,a_{s_1}\T x$
or equivalently 
\[
c_0(a_{s_0}\T y-a_{s_0}\T x)= c_1(a_{s_1}\T x-a_{s_1}\T y),
\]
so $c_0$ and $c_1$ have the same sign because $x$ is not on
facet $F_{s_0}$ and $y$ is not on facet $F_{s_1}$, so
$a_{s_0}\T y-a_{s_0}\T x= b_{s_0}-a_{s_0}\T x>0$ 
and $a_{s_1}\T x-a_{s_1}\T y= b_{s_1}-a_{s_1}\T y>0$.
By (\ref{lindep}),
\[
0=\det[(a_{s_0}c_0+a_{s_1}c_1)~a_{s_2}\cdots a_{s_m}]
=c_0\det[a_{s_0}a_{s_2}\cdots a_{s_m}]
+c_1\det[a_{s_1}a_{s_2}\cdots a_{s_m}]
\]
which shows that 
$\det[a_{s_0}a_{s_2}\cdots a_{s_m}]$ and
$\det[a_{s_1}a_{s_2}\cdots a_{s_m}]$ have indeed opposite sign.
\endproof

The orientation of a vertex of a simple polytope $P$ depends
only on the determinant of the normal vectors $a_j$ of the
facets in (\ref{det}), but not on the right hand sides $b_j$
when $P$ is given as in (\ref{P}).
Translating the polytope $P$ by adding a constant vector to
each point of $P$ only changes these right hand sides.
If $\0$ is in the interior of~$P$, then one can assume that
$b_j=1$ for all $j$ in~$[n]$.
The convex hull of the vectors $a_j$ is then a simplicial
polytope $P^\Delta$ called the ``polar'' of $P$ (see
Ziegler, 1995).
The vertices of $P^\Delta$ correspond to the facets of~$P$
and vice versa.
A pivoting system for the simplicial polytope has its
vertices as nodes and its facets as states, which one may
see as a more natural definition.
However, the facets of a simplicial polytope are oriented via
(\ref{det}) only if it has $\0$ in its interior, which is
not required for the simple polytope~$P$.
For common descriptions such as (\ref{Pge0}), we therefore
prefer to look at simple polytopes.

Theorem~\ref{t-sign} and Proposition~\ref{p-det} replicate,
in streamlined form, Shapley's (1974) proof that the
equilibria at the ends of a Lemke--Howson path have opposite
index.
Applied to the polytope $P$ in (\ref{Pge0}), the completely
labeled vertex $\0$ does not represent a Nash equilibrium,
and it is customarily assumed to have index $-1$, which is
achieved by multiplying all orientations with $-1$ if $m$ is
even.

\section{Oriented Euler complexes}
\label{s-oiks}

Todd (1972; 1974) introduced the concept of a
``semi-duoid'', which was studied by Edmonds (2009) under
the name of Euler complex or ``oik''.
Edmonds showed that ``room partitions'' for a ``family of
oiks'' come in pairs.
In this section, we give a direction to Edmonds's parity
argument.
For that purpose, we introduce the new concept of an
{\em oriented oik} and show that one can then define signs
for ``ordered room partitions'', where the order of the
rooms can be disregarded for oiks of even dimension
(Theorem~\ref{t-part}).
      We discuss the connection of labels with ``Sperner
      oiks'' in Appendix~A.

\begin{definition}
\label{d-oik}
Let $V$ be a finite set of {\em nodes} and let $d$ be an
integer, $d\ge2$.
A $d$-dimen\-sio\-nal {\em Euler complex} or {\em $d$-oik} on
$V$ is a multiset $\rooms$ of $d$-ele\-ment subsets of $V$,
called {\em rooms}, so that any set $W$ of $d-1$ nodes is
contained in an even number of rooms.
If $W$ is always contained in zero or two rooms, then the
oik is called a {\em manifold}.
A {\em wall} is a $(d-1)$-element subset of a room $R$.  
A {\em neighboring} room to $R$ for a wall $W$ of $R$ is any
room that contains $W$ as a subset.
\end{definition}

In the preceding definition we follow Edmonds, Gaubert, and
Gurvich (2010) of choosing $d$ rather than $d-1$ (as in
Edmonds, 2009) for the dimension of the oik.
A 2-oik on $V$ is an Euler graph with node set $V$ and edge
multiset $\rooms$.
We allow for parallel edges (which is why $\rooms$ in
Definition~\ref{d-oik} is a multiset, not a set) but no
loops.

Rooms are often called ``abstract simplices'', and a longer
term for manifold is ``abstract simplicial pseudo-manifold''
(e.g., Lemke and Grotzinger, 1976).
The following definition generalizes the common definition
of coherently oriented rooms in manifolds (Hilton and Wylie,
1967, p.~54) to oiks.

\begin{definition}
\label{d-oikorient}
Consider a $d$-oik $\rooms$ on $V$ and fix a linear order
on~$V$.
Represent each room $R=\{s_1,\ldots,s_d\}$ in $\rooms$ as
$r(R)=(s_1,\ldots,s_d)$ where $s_1,\ldots,s_d$ are in
increasing order.
For each room~$R$, choose an {\em orientation} $\sigma(R)$
in $\{-1,1\}$.
The {\em induced orientation} on any wall $W=R-\{s_i\}$ is
defined as $(-1)^i\sigma(R)$.
The orientation of the rooms is called {\em coherent}, and
the oik {\em oriented}, if half of the rooms containing any
wall $W$ induce orientation $1$ on $W$ and the other half
orientation~$-1$ on~$W$.
\end{definition}

As an example, consider a 2-oik, where rooms are the edges
of an Euler graph.
Suppose an edge $\{u,v\}$ is oriented so that $\sigma(u,v)=1$.
Then the induced orientation on the wall $\{u\}$ is $-1$ and
on $\{v\}$ it is $1$, so $\{u,v\}$ becomes the edge $(u,v)$
of a digraph oriented from $u$ to~$v$.
A coherent orientation means that each wall (that is, node)
has as many incoming as outgoing edges, so this is an
Eulerian orientation of the graph (which always exists; for
$d>2$ there are already manifolds that cannot be oriented,
for example a triangulated Klein bottle).
In general, the simplest oriented oik consists of just two
rooms with equal node set but opposite orientation.
As an Euler digraph, this is a pair of oppositely
oriented parallel edges.

\begin{proposition}
\label{p-oikpivot}
A $d$-oik $\rooms$ on $V$ defines a pivoting system
$(S,V,m,r,f)$ as follows:
Let $S=\rooms$, $m=d$, 
and $r$ and $\sigma$ be as in Definition~\ref{d-oikorient}.
For any wall $W$, match the $2k$ rooms that contain $W$ into
$k$ pairs $(R,R')$, where $R$ and $R'$ induce opposite
orientation on $W$ if the oik is oriented. 
Then $f(R,i)=R'$ if $r(R)=(s_1,\ldots,s_d)$ and
$W=R-\{s_i\}$.
If $\sigma$ is coherent, then the pivoting system is
oriented.
\end{proposition}

\proof
Let $R\cup
R'=\{s_1,\ldots,s_{d+1}\}=R\cup\{s_j\}=R'\cup\{s_i\}$,
with $s_1,\ldots,s_{d+1}$ in increasing order, and let $i<j$,
otherwise exchange $R$ and $R'$.
Then $r(R')$ is obtained from $r(R)$ by replacing $s_i$ with
$s_j$ followed by the permutation $\pi$ that inserts
$s_j$ at its place in the ordered sequence by ``jumping
over'' $j-i-1$ elements $s_{i+1},\ldots,s_{j-1}$ to remove
as many inversions, so $\parity(\pi)=(-1)^{j-i-1}$.
Hence, $f(R,i)=R'$ is well defined.
If $\sigma$ is coherent, then $R$ and $R'$ induce on the
common wall $R\cap R'$ the opposite orientations
$(-1)^i\sigma(R)$ and $(-1)^{j-1}\sigma(R')$
(because $s_i\not\in R'$), that is,
$\sigma(R')=-\sigma(R)(-1)^{j-i-1}=-\sigma(R)\cdot\parity(\pi)$
as required in (\ref{switch}).  
\endproof

The matching of rooms with a common wall into $k$ pairs
described in Proposition~\ref{p-oikpivot} is unique if the
oik is a manifold.
In a 2-oik, that is, an Euler graph, such a matching of
incoming and outgoing edges of a node is for example obtained
from an Eulerian tour of the graph, which also gives a
coherent orientation.

For an ``oik-family'' $\rooms_1,\ldots,\rooms_h$ where each
$\rooms_p$ is a $d_p$-oik on the same node set $V$ for
$p\in[h]$, Edmonds, Gaubert, and Gurvich (2010) define the
``oik-sum'' as follows.

\begin{definition}
\label{d-oiksum}
Let $\rooms_p$ be a $d_p$-oik on $V$ for $p\in[h]$, and let
$m=\sum_{p=1}^h d_p$.
Then the {\em oik-sum} $\rooms=\rooms_1+\cdots+\rooms_h$
is defined as the set of $m$-element subsets $R$ of
$[h]\times V$ so that
\begin{equation}
\label{uplus}
R=R_1\uplus R_2\uplus\cdots\uplus R_h=
(\{1\}\times R_1)\cup(\{2\}\times R_2)\cup\cdots\cup(\{h\}\times R_h)
\end{equation}
where $R_p\in\rooms_p$ for $p\in[h]$.
For a fixed order $<$ on $V$, we order $[h]\times V$
lexicographically by $(p,u)<(q,v)$ if and only if $p<q$,
or $p=q$ and $u<v$.
\end{definition}

As observed by Edmonds, Gaubert, and Gurvich (2010),
the oik-sum $\rooms$ is an oik.
A neighboring room of $R=R_1\uplus R_2\uplus\cdots\uplus R_h$
is obtained by replacing, for some $p$, the room $R_p$ with 
a neighboring room $R_p'$ in $\rooms_p$.
The next proposition states, as a new result, that the
oik-sum is oriented if each $\rooms_p$ is oriented.
According to Definition~\ref{d-oikorient}, this requires an
order on the node set $[h]\times V$ to yield an order on the
nodes in room $R$ in (\ref{uplus}), which is provided in
Definition~\ref{d-oiksum}:
The nodes of each room $R_p$ are
listed in increasing order (on~$V$), and these $d_p$-tuples
are then listed in the order of the rooms $R_1,\ldots,R_h$;
this becomes the representation $r(R)$ used to define the
orientation $\sigma$ on~$\rooms$.

\begin{proposition}
\label{p-sumoik}
The oik-sum $\rooms$ in Definition~\ref{d-oiksum} is an
$m$-oik over $[h]\times V$.
If each $\rooms_p$ is oriented with $\sigma_p$, so is $\rooms$, with
\begin{equation}
\label{sumorient}
\sigma(R_1\uplus\cdots\uplus R_h)=\prod_{p=1}^h\sigma_p(R_p).
\end{equation}
\end{proposition}

\proof
Clearly, each room $R$ of $\rooms$ as in (\ref{uplus}) has
$m$ elements.  
Any wall $W$ of $R$ is given by $W=R-\{(p,v)\}$ for some
$p$ in $[h]$ and $v$ in $R_p$.
Then any neighboring room $R'$ in $\rooms$ of $R$ for the
wall $W$ is given by 
\[
R'=
R_1\uplus\cdots\uplus R_{p-1}\uplus R'_p\uplus R_{p+1}\cdots\uplus R_{h}
\]
for the neighboring rooms $R'_p$ in $\rooms_p$ for $R_p-\{v\}$,
of which, including $R_p$, there is an even number.
This shows that $\rooms$ is an $m$-oik.

For the orientation of $\rooms$ if each $\rooms_p$ is
oriented with $\sigma_p$, represent $R$ as $r(R)$ by listing
the elements of $R$ in lexicographic order as in
Definition~\ref{d-oiksum}.
Then the induced orientation on any wall $W=R-\{(p,v)\}$ as
in Definition~\ref{d-oikorient} is obtained from the induced
orientation on $R_p-\{v\}$, as follows.
Suppose $s^p_1,\ldots,s^p_{d_p}$ are the nodes in $R_p$ in
increasing order, where $v=s^p_i$.
Then the induced orientation on $R_p-\{v\}$ in $\rooms_p$ is
$(-1)^i\sigma_p(R_p)$.
In $r(R)$, node $v$ appears in position
$\sum_{j=1}^{p-1}d_j+i$, so the induced orientation of $R$
on $W$ is, with $\sigma(R)$ is defined as in (\ref{sumorient}),  
\begin{equation}
\label{induced}
(-1)^{\sum_{j=1}^{p-1}d_j+i}\sigma(R)=
(-1)^i\sigma_p(R_p)
(-1)^{\sum_{j=1}^{p-1}d_j}
\prod_{q\in[h]-p}\sigma_q(R_q).
\end{equation}
All the rooms in $\rooms$ that contain $W$ are obtained by
replacing $R_p$ with any room $R'_p$ that contains
$R_p-\{v\}$.
Half of these have induce the same orientation as $R_p$ on
$R_p-\{v\}$, half of these the other orientation.
Because this affects only the term $(-1)^i\sigma_p(R_p)$ in
(\ref{induced}), half of the rooms $R'$ that contain $W$
induce one orientation on $W$ and half the other
orientation.
So $\sigma$ is a coherent orientation of $\rooms$.
\endproof

Consider now an oik-family $\rooms_1,\ldots,\rooms_h$ where 
$\rooms_p$ is a $d_p$-oik on $V$ for
$p$ in $[h]$ so that $|V|=m=\sum_{p=1}^h d_p$.
Suppose $R_p\in\rooms_p$ for $p$ in $[h]$ and
$\bigcup_{p=1}^hR_p=V$ (so the rooms $R_p$ are, as subsets
of $V$, also pairwise disjoint).
Then $(R_1,\ldots,R_h)$ is called an {\em ordered room
partition}.
In the following theorem, the even number of ordered room
partitions is due to Edmonds, Gaubert, and Gurvich (2010);
the observation on signs is new.  

\begin{theorem}
\label{t-ordpart}
Let $\rooms_p$ be a $d_p$-oik on $V$ for
$p$ in $[h]$ and $|V|=m=\sum_{p=1}^h d_p$.
Then the number of ordered room partitions is even.
If each $\rooms_p$ is oriented as in
Proposition~\ref{p-sumoik}, then there is an equal
number of ordered room partitions of positive as of negative
sign, where the sign of a room partition
$(R_1,\ldots,R_h)$  is defined by 
\begin{equation}
\label{partsign}
\sign(R)=\sign(R_1,\ldots,R_h)=
\sigma(R_1\uplus\ldots\uplus R_h)\cdot\parity(\pi)
\end{equation}
with the permutation $\pi$ of $V$ given according to the
order of the nodes of $V$ in $r(R)$, that is, with 
$\pi(u)<\pi(v)$ if $u\in R_p$ and $v\in R_q$ and $p<q$, 
or $u,v\in R_p$ and $u<v$ in~$V$.
\end{theorem}

\proof
This is a corollary of Theorem \ref{t-sign} and
Propositions \ref{p-oikpivot} and \ref{p-sumoik}. 
Assume that $V=\{v_1,\ldots,v_m\}$ with the order on $V$
given by $v_i<v_j$ for $i<j$ (or just let $V=[m]$).
Define the labeling $l:[h]\times V\to[m]$ by $l(p,v_i)=i$
for $i\in[m]$.
Then the CL rooms $R_1\uplus\ldots\uplus R_h$ of
$\rooms_1+\cdots+\rooms_h$ are exactly the ordered room
partitions, with the sign in (\ref{partsign}) defined as in
(\ref{CLsign}).
So there is an equal number of them of either sign.

If the oiks are not all oriented, then the paths that
connect any two CL states are still defined, so the number
of ordered room partitions is even, except that they have no
well-defined sign.
\endproof

Connecting any two room partitions by paths of ACL states as
in the preceding proof corresponds to the ``exchange graph''
argument of Edmonds (2009), where the ACL states correspond
to {\em skew room partitions} $(R_1,\ldots,R_h)$ defined by
the property $\bigcup_{p=1}^hR_p=V-\{w\}$ for some $w$
in~$V$; here $w$ represents the missing label.

Suppose now that all oiks $\rooms_p$ in the oik family are
the same $d$-oik $\rooms'$ over $V$ for $p$ in $[h]$, with
$|V|=m=h\cdot d$.
Then any ordered room partition $(R_1,\ldots,R_h)$ defines
an (unordered) {\em room partition} $\{R_1,\ldots,R_h\}$.
Any such partition gives rise to $h!$ ordered room
partitions, so if $h\ge2$ their number is trivially even.
However, the path-following argument can be applied to 
the unordered partitions as well (which is the original
exchange algorithm of Edmonds, 2009), which shows that the
ordered room partitions at the two ends of the pivoting path
define different unordered partitions.
The next theorem shows that unordered partitions
$\{R_1,\ldots,R_h\}$ are connected by pivoting paths, which
are essentially the same paths as in
Theorem~\ref{t-ordpart}, and that the sign property
continues to hold when $d$ is even and $\rooms'$ is
oriented.

\begin{theorem}
\label{t-part}
Let $\rooms'$ be a $d$-oik on $V$ and $|V|=m=h\cdot d$.
Then the number of room partitions $\{R_1,\ldots,R_h\}$
is even.
If $\rooms'$ is oriented with $\sigma'$and $d$ is even, then
$\sign(R_1,\ldots,R_h)$ as defined in $(\ref{sumorient})$
with $\sigma_p=\sigma'$ and $(\ref{partsign})$ is
independent of the order of the rooms $R_1,\ldots,R_h$, and
there are as many room partitions of sign $1$ as of
sign~$-1$.
\end{theorem}

\proof
We consider unordered multisets $\{R_1,\ldots,R_h\}$ of $h$
rooms of $\rooms'$ as states $s$ of a pivoting system.  
We first define a representation $r(s)=(s_1,\ldots,s_m)$.
Let $R_p=\{s^p_1,\ldots,s^p_d\}$ for $p$ in $[h]$
where $s^p_1,\ldots,s^p_d$ are in increasing order according
to the order on $V$.
Fix some order of the rooms in $\rooms'$, for example
the lexicographic order with some tie-breaking for rooms
that have the same node set.
Assume that the rooms $R_1,\ldots,R_h$ are in ascending
order, which defines a unique representation of $s$ as 
\begin{equation}
\label{sort}
r(s)=r(\{R_1,\ldots,R_h\})=
(s^1_1,\ldots,s^1_d, ~
s^2_1,\ldots,s^2_d,~
\ldots, 
s^h_1,\ldots,s^h_d).
\end{equation}
(Note that $r$ may not be injective, which is allowed.)
Assume that neighboring rooms in $\rooms'$ are matched into
pairs $R_p,R_p'$ containing the wall $R_p-\{s_i\}$ as in
Proposition~\ref{p-oikpivot}.
The pivoting step from $s=\{R_1,\ldots,R_h\}$ to $t=f(s,i)$
replaces $R_p$ by $R'_p$.

In (\ref{sort}), the nodes of each individual room $R_p$
still appear consecutively as in the permutation $\pi$ in
Theorem~\ref{t-ordpart}, except for the order of the rooms
themselves.
Then with $v_1,\ldots,v_m$ as the nodes of $V$ in increasing
order and the ``identity'' labeling $l:V\to[m]$, $l(v_i)=i$,
the $m$-tuple $l(r(s))$ defines a
permutation $\pi$ of $[m]$ if $s$ is a room
partition, as in (\ref{partsign}).
Then the parity of $\pi$ does not depend on the order of the
rooms in $s$ if $d$ is even, so the sign in (\ref{partsign})
is well defined and the same as in (\ref{CLsign}).
An ACL state $s$ is a skew room partition,
which has two opposite signs as in (\ref{opposite}).
Then the claim follows from Theorem~\ref{t-sign}.
\endproof

\begin{figure}[hbt] 
\strut\hfill
\epsfxsize5cm
\epsfbox{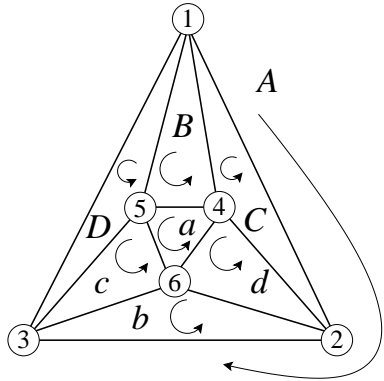} 
\hfill
\strut 
\caption{A 3-oik with triangles as rooms.
The circular arrows indicate the positive orientation of nodes
in a room.
} 
\label{octahedron}
\end{figure} 

The following example shows that we cannot expect to define
a sign to unordered room partitions when $\rooms'$ has odd
dimension~$d$ (see also Merschen, 2012, Figure~3.6).
Let $d=3$ and consider the oik defined by the eight vertices
of the 3-dimensional cube, which correspond to the facets of
the octahedron, shown as the triangles in
Figure~\ref{octahedron} including the outer triangle
marked~``$A$''.
A coherent orientation of the eight rooms is obtained as
follows (shown in Figure~\ref{octahedron} with a circular
arrow that shows the positively oriented order of the
nodes): 
$\sigma(A)=\sigma(123)=1$,
$\sigma(B)=\sigma(145)=-1$,
$\sigma(C)=\sigma(124)=-1$,
$\sigma(D)=\sigma(135)=1$,
$\sigma(a)=\sigma(456)=1$,
$\sigma(b)=\sigma(236)=-1$,
$\sigma(c)=\sigma(356)=-1$,
$\sigma(d)=\sigma(246)=1$.
The four room partitions are
$\{A,a\}$,
$\{B,b\}$,
$\{C,c\}$,
$\{D,d\}$.
Any two of these are connected by pivoting paths, so they
cannot always have opposite signs at the end of these paths.
However, for ordered room partitions the signs work.
For example, $(A,a)$ is connected to $(b,B)$ via the
complementary pivoting steps
$(123,456)\to(236,456)\to(236,145)$, and to $(C,c)$ via the
steps
$(123,456)\to(124,456)\to(124,356)$.
Moreover, $(C,c)$ connects to $(B,b)$ via
$(124,356)\to(145,356)\to(145,236)$.
We have $\sign(A,a)=1$, $\sign(b,B)=-1$ (because $236\,145$
has parity $-1$), and $\sign(C,c)=-1$ and $\sign(B,b)=1$.
The two ordered room partitions $(b,B)$ and $(B,b)$ have
different signs because they define two permutations
$236\,145$ and $145\,236$ of opposite parity.

\section{Related work}
\label{s-related} 

Todd (1972; 1974; 1976) developed an abstract theory of
complementary pivoting, using
``semi-primoids'' and ``semi-duoids''.
A semi-duoid is the same as an ``oik'' as defined by
Edmonds (2009), see Definition~\ref{d-oik} above.
For a semi-duoid $\rooms$ on $V$, the set $\{V-R\mid
R\in\rooms\}$ is a semi-primoid.
(``Primoids'' and ``duoids'' fulfill an additional
connectness condition.)
For example, for the basic feasible solutions of a system
of linear equations with nonnegative variables, the sets of
basic variables form a primoid and the sets of nonbasic
variables form a duoid.
% % a ``manifold'' called ``proper semi-duoid'' by Todd.
% Todd (1972; 1974; 1976) developed an abstract theory of
% complementary pivoting.
% He defined ``semi-primoids'' and ``semi-duoids'' which are
% sets of subsets of a finite set $V$ with certain properties
% (``primoids'' and ``duoids'' fulfill an additional
% connectivity condition).
% For example, for the basic feasible solutions of a system
% of linear equations with nonnegative variables, the sets of
% basic variables form a primoid and the sets of nonbasic
% variables form a duoid.
% A semi-duoid is the same as an ``oik'' as defined by
% Edmonds (2009), see Definition~\ref{d-oik} above.
% For a semi-primoid $\rooms$ on $V$, the set $\{V-R\mid
% R\in\rooms\}$ is a semi-duoid.

Todd defines the pivoting operation by alternating
between the semi-duoid and the semi-primoid.
Edmonds defines pivoting by exchanging a room with an
adjacent room.
Edmonds shows that partitions of $V$ into rooms for a given
``oik family'' come in pairs.
This result is equivalent to that of Todd for partitions of
$V$ into two rooms, but more general when considering
partitions into more than two rooms.
In order to obtain a unique path of complementary pivoting,
Todd (1974, p.~255) describes the local pairing of the
$2k$ rooms that contain a common wall into $k$ pairs as in
Proposition~\ref{p-oikpivot}.
In contrast, Edmonds (2009, p.~66) merely stipulates ``no
repetition'' which requires remembering the history of the
pivoting path.

The Lemke--Howson algorithm finds a Nash equilibrium of an
$m\times n$ bimatrix game.
Its pivoting steps alternate between vertices of two
polytopes of dimension $m$ and $n$, respectively (see von
Stengel, 2002, for an exposition).
In order to capture this with room partitions, Edmonds
(2009) considers two oiks of (possibly different) dimension
$m$ and $n$, respectively, on the same set $V=[m+n]$.
However, alternating between two polytopes is not essential,
by considering instead their product as a single labeled
polytope, as described in Section~\ref{s-labpoly} above.

We have described complementary pivoting using labels, with
the pivoting step started by the missing label and on the
path determined by the duplicate label.
A given labeling (or ``coloring'') $V\to[m]$ determines an
oik $\rooms_0$ of dimension $|V|-m$ whose elements are the
complements of completely labeled sets.
It is a manifold (also known as the ``coloring manifold'')
where removing any node $u$ from a room $R$ and replacing it
with the unique node in $V-R$ with the same label as $u$
gives the adjacent room.  
If $\rooms$ is an $m$-oik on $V$, then the completely
labeled rooms $R$ of $\rooms$ are clearly those so that
$\{R,R_0\}$ with $R_0\in\rooms_0$ is a partition of~$V$.
Edmonds, Gaubert, and Gurvich (2010) call $\rooms_0$ a
``Sperner oik''.
The oik $\rooms_0$ is ``polytopal'' because its rooms
correspond to the vertices of a product of simplices
(Edmonds, 2009, Example~3).
This has also been observed by Todd (1972, p.~1.5; 1974,
p.~248) who calls $\rooms_0$ a ``simplicial duoid''.
A similar product of simplices results from the constraints
$y\ge\0$, $Ay\le\1$ in (\ref{AyCx}) for the unit-vector game
$(A,C\T)$ in Proposition~\ref{p-unitv}, where each column of
$A$ is a unit vector.

Edmonds, Gaubert, and Gurvich (2010) show that the pivoting
path for a family of oiks on $V$ can instead be applied to
room partitions for only two oiks, namely their oik-sum (see
Definition~\ref{d-oiksum} above) together with a Sperner oik
$\rooms_0$.
Oik-sums are equivalent to products of semi-duoids defined
by Todd (1972, Chapter~5).
This seems to reduce everthing to Todd (1972) who covered
the case of two oiks.
However, as already mentioned, partitions of $V$ into more
than two rooms (even if implied by a suitable oik-sum) were
not explicitly considered by Todd.

The labels used by Edmonds, Gaubert, and Gurvich (2010) to
define the Sperner oik $\rooms_0$ are the elements of~$V$.
This is essentially the same argument as our proof of
Theorem~\ref{t-ordpart}, without the orientation.
%In V\'egh and von Stengel (2012, Appendix~A), we argue that
In Appendix~A, we argue that
the definition of a ``sign'' requires a reference to the
parity of the permutation of the labels of a room, which
does not seem simpler when looking at room partitions with a
Sperner oik instead.

Shapley (1974) showed that the Nash equilibria at the two
ends of a Lemke--Howson path have opposite index, defined in
terms of determinants of the payoff matrices restricted to
the equilibrium support.
(That paper also gives an accessible exposition of the
Lemke--Howson algorithm using ``labels''.) 
Theorem~\ref{t-sign} and Proposition~\ref{p-det} replicate
Shapley's argument in streamlined form.
Lemke and Grotzinger (1976) define coherent orientations of 
abstract simplicial manifolds.
Our approach is similar, except that we separate states and
their representation, and apply the concepts of orientation
and sign to the representation, in order to capture room
partitions as well, and oiks that are not manifolds.
If the oik cannot be oriented, then Lemke and Grotzinger
(1976) have shown (for nonorientable manifolds) that
opposite signs for CL rooms cannot be defined in general;
see also Grigni (2001).

Todd (1976) extends the alternate primoid-duoid pivoting
steps with an orientation, and also simplifies Shapley's
approach.
His construction is essentially equivalent to that of Lemke
and Grotzinger (1976).
It also does not extend to room partitions with more than
two rooms, nor to oiks that are not manifolds (Todd, 1976,
p.~54).
Our own contribution is a framework of {\em oriented}
complementary pivoting that encompasses room partitions in
oiks, for which orientations are new.

Eaves and Scarf (1976, Sections 5--6) apply index theory to
piecewise linear mappings in a more general setting, which
we have not tried to include in our model.

One of our main examples of partitions of $V$ into more
than two rooms is perfect matchings in an Euler graph as
considered by Edmonds (2009, Example~4) (but, to our
knowledge, not by Todd or others).
For an Euler digraph, these perfect matchings have a sign,
which has been studied in the context of Pfaffian
orientations of a graph; we discuss this connection in
Section~\ref{s-matchings} to keep that section largely
self-contained.

Interestingly, perfect matchings of an Euler digraph
correspond to CL vertices of a labelled ``dual cyclic
polytope''.
These polytopes have been used by Morris (1994) to construct
exponentially long Lemke paths, and by Savani and von
Stengel (2006) to construct exponentially long
Lemke--Howson paths.
The connection to Euler digraphs is due to Casetti,
Merschen, and von Stengel (2010) and Merschen (2012) and is
summarized at the end of Section~\ref{s-matchings}.  

\section{Signed perfect matchings}
\label{s-matchings} 

This section is concerned with algorithmic questions of room
partitions in 2-oiks, which are perfect matchings in Euler
graphs.
The sign of a perfect matching, for any orientation of the
edges of a graph, is closely related to the concept of a
{\em Pfaffian orientation} of a graph, where all perfect
matchings have the same sign.
The computational complexity of finding such an orientation
is an open problem (see Thomas, 2006, for a survey).
An Eulerian orientation is not Pfaffian by
Theorem~\ref{t-part}, a fact that is also easy to verify
directly.
The main result of this section (Theorem~\ref{t-euler})
states that in an Euler digraph, a
second perfect matching of opposite sign can be found in
polynomial (in fact, near-linear) time.
This holds in contrast to the complementary pivoting
algorithm, which can take exponential time; Casetti,
Merschen, and von Stengel (2010) have shown how to apply
results of Morris (1994) for this purpose.
However, the pivoting algorithm takes linear time in a {\em
bipartite} Euler graph, and a variant can be used to find an
oppositely signed matching in a bipartite graph that has no
source or sink (Proposition~\ref{p-bip}).

We follow the exposition of Pfaffians in Lov\'asz and
Plummer (1986, Chapter~8).
The determinant of an $m\times m$ matrix $B$ with entries
$b_{ij}$ is defined as
\begin{equation}
\label{defdet}
\det \,B=\sum_\pi \parity(\pi) \prod_{i=1}^m b_{i,\pi(i)}
\end{equation}
where the sum is taken over all permutations $\pi$ of $[m]$.
Let $B$ be skew symmetric, that is, $B=-B\T$.
Then $\det\,B=\det(-B\T)=\det(-B)=(-1)^m\det\,B$, so 
$\det\,B=0$ if $m$ is odd. 
Assume $m$ is even.
Then it has long been known (see references below) that
% $\det\,B$ can be written as the square of a function
% $\pf B$ called the {\em Pfaffian} of $B$,
%The square of the Pfaffian is the determinant,
\begin{equation}
\label{pf}
\det\, B=(\pf B)^2
\end{equation}
for a function $\pf B$ called the {\em Pfaffian} of $B$,
defined as follows.
Let $\match(m)$ be the set of all partitions $s$ of $[m]$
into pairs, $s=\{\{s_1,s_2\},\ldots,\{s_{m-1},s_m\}\}$, and
let $\parity(s)$ be the parity of $(s_1,s_2,\ldots,s_m)$
seen as a permutation of $[m]$ under the assumption that
each pair $\{s_{2k-1},s_{2k}\}$ is written in increasing
order, that is, $s_{2k-1}<s_{2k}$ for $k$ in $[m/2]$; the
order of the pairs themselves does not matter.
Then 
%the {\em Pfaffian} of $B$ is defined as
\begin{equation}
\label{defpf}
\pf B=\sum_{s\in\match(m)}\parity(s)
\prod_{k=1}^{m/2}b_{s_{2k-1},\,s_{2k}}~.
\end{equation}
In fact, because $B$ is skew symmetric, the order % of nodes
of a pair $(s_{2k-1},s_{2k})$ can also be changed because
this also changes the parity of $s$.
An example of (\ref{defpf}) is $m=4$ where
% For example, if $m=4$, then 
$\pf B=b_{12}b_{34}-b_{13}b_{24}+b_{14}b_{23}$.

% This theorem is attributed to Cayley and a standard
% reference is to Muir (1960) which we found impenetrable;
Parameswaran (1954) and Lax (2007, Appendix~2) show that a
skew-symmetric matrix $B$ fulfills (\ref{pf}) for some
function $\pf B$.
% without its explicit form~(\ref{defpf}).
For a direct combinatorial proof, one can see that the
products in (\ref{defdet}) are zero for those permutations
$\pi$ where $\pi(k)=k$ for some~$k$, and cancel out for the
permutations with odd cycles;
then only permutations with even-length cycles remain, which
can be obtained uniquely, using those cycles, from pairs of
partitions taken from $\match(m)$ (see also Jacobi, 1827,
pp.\ 354ff, and Cayley, 1849).

Consider a simple graph $G$ with node set $[m]$.
An {\em orientation} of $G$ creates a digraph by giving
each edge $\{u,v\}$ an orientation as $(u,v)$ or $(v,u)$.
Define the $m\times m$ matrix $B$ via
\begin{equation}
\label{edge}
b_{uv}=\left\{\begin{array}{rl}
0~~&
\hbox{if $\{u,v\}$ is not an edge,}\\
1~~ & 
\hbox{if $\{u,v\}$ is oriented as $(u,v)$,}\\
-1~~ & 
\hbox{if $\{u,v\}$ is oriented as $(v,u)$.}\\
\end{array}
\right.
\end{equation}
Then $B$ is skew symmetric.
Any $s$ in $\match(m)$ is a perfect matching of $G$ if and
only if $\prod_{k=1}^{m/2}b_{s_{2k-1},\,s_{2k}}\ne0$, so
only the perfect matchings of $G$ contribute to the sum
in~(\ref{defpf}).

If $G$ is an Euler digraph, that is, an oriented 2-oik, then
this defines the orientation of edge $\{u,v\}$, assuming
$u<v$, as $\sigma(\{u,v\})=b_{uv}$, according to
Definition~\ref{d-oikorient}.
Then by (\ref{sumorient}) and (\ref{partsign}), a perfect
matching $s$ has the sign 
\[
\sign(s)=\parity(s_1,\ldots,s_m)\cdot
\prod_{k=1}^{m/2}b_{s_{2k-1},\,s_{2k}}~,
\]
so the Pfaffian $\pf B$ in (\ref{defpf}) is the sum over all
matchings of $G$ weighted with their signs.
For the Eulerian orientation, that sum is zero by
Theorem~\ref{t-part}, which follows also from (\ref{pf})
because $B\1=\0$, so $\det\,B=0$.

In our Definition~\ref{d-oik} of a $d$-oik, $\rooms$ can be a
multiset, which for $d=2$ defines an Euler graph $G$ which
may have parallel edges and then is not simple.
The rooms themselves have to be sets, so loops are not
allowed.
In this case, (\ref{edge}) can be extended to define
$b_{uv}$ as the number of edges oriented as $(u,v)$ minus
the number of edges oriented as $(v,u)$.
This counts the number of matchings with their signs
correctly; oppositely oriented parallel edges $(u,v)$ and
$(v,u)$ cancel out both in contributing to $b_{uv}$ and
when counting matchings with their signs.

For any graph $G$ and any orientation of $G$, the sign of a
perfect matching $s$ is most easily defined by writing down
the nodes of each edge $\{s_{2k-1}, s_{2k}\}$ in the way the
edge is oriented as $(s_{2k-1}, s_{2k})$;
this does not affect (\ref{defpf}) as remarked there.
When writing down the nodes $s_1,\ldots,s_m$ this way,
$\sign(s)=\parity(s_1,\ldots,s_m)$ and 
$\pf B=\sum_{s\in\match(G)}\sign(s)$ where $\match(G)$ is
the set of perfect matchings of~$G$.

A {\em Pfaffian orientation} is an orientation of $G$ so
that all perfect matchings have positive sign.  
Its great computational advantage is that it allows to 
compute the number of perfect matchings of $G$ using
(\ref{pf}) by evaluating the determinant $\det\,B$, which
can be done in polynomial time.
In general, counting the number of perfect matchings is
\#P-hard already for bipartite graphs (Valiant, 1979).
The question if a graph has a Pfaffian orientation is
polynomial-time equivalent to deciding whether a given
orientation is Pfaffian (see Vazirani and Yannakakis, 1989,
and Thomas, 2006).
For bipartite graphs, this problem is equivalent to finding
an even-length cycle in a digraph, which was long open and
shown to be polynomial by Robertson, Seymour, and Thomas
(1999).
For general graphs, its complexity is still open.

We now consider the following algorithmic problem: 
Given an Euler digraph with a perfect matching, find another
matching of opposite sign, which exists.
Without the sign property, a second matching can be found by
removing one of the given matched edges from the graph and
applying the ``blossom'' algorithm of Edmonds (1965) to find 
a maximum matching, which finds another perfect matching for
at least one removed edge; however, its sign cannot be
predicted, and adapting this method to account for the sign
seems to lead to the difficulties related to Pfaffian
orientations in general graphs.
Merschen (2012, Theorem~5.3) has shown how to find in
polynomial time an oppositely signed matching in a planar
Euler graph, and his method can be adapted to graphs that,
like planar graphs, are known to have a Pfaffian orientation.

The following theorem presents a surprisingly
simple algorithm for any Euler graph.
It runs in near-linear time in the number of edges of the
graph and is faster and simpler than using blossoms.
The inverse Ackermann function $\alpha$ is an extremely
slowly growing function with $\alpha(n)\le 4$ for
$n\le2^{2048}$ (Cormen et al., 2001, Section 21.4).

\begin{theorem}
\label{t-euler}
Let $G=(V,E)$ be an Euler digraph, and let $M$ be a perfect
matching of $G$.
Then a perfect matching $M'$ of opposite sign can be found
in time $O(|E|\cdot\alpha(|V|))$, where $\alpha$ is the inverse
Ackermann function.
\end{theorem}

\proof
The matching $M$ is a subset of $E$.
A {\em sign-switching cycle} $C$ is an even-length cycle 
so that every other edge in $C$ belongs to $M$,
and so that, in a chosen direction of the cycle, $C$ has an
even number of forward-oriented edges. 
We claim that then the symmetric difference
$M'=M\triangle C$ has opposite sign to~$M$.
To see this, suppose first that all edges in~$C$ point
forward, and that $C\cap M$ consists of the first $k/2$
edges $(s_1,s_2)$, \ldots, $(s_{k-1},s_k)$ of~$M$ (which
does not affect the sign of~$M$).
Then these edges are replaced in $M'$ by
$(s_k,s_1)$, $(s_2,s_3)$, \ldots, $(s_{k-2},s_{k-1})$,
which defines an odd permutation of these $k$ nodes, so
$M'$ has opposite sign to~$M$.
Changing the orientation of any two edges in $C$ leaves the
sign of both $M$ and $M'$ unchanged (if both edges belong to
$M$ or to~$M'$) or changes the signs of both $M$ and $M'$,
so they stay opposite.
This proves the claim.

So it suffices to find a sign-switching cycle $C$ for $M$,
which is achieved by the following algorithm: Successively
apply one of the following reductions (a) or (b) to $G$
until (c) applies:

\abs{(a)}
If $v$ in $V$ has indegree and outdegree $1$ with edges 
$(u,v)$ and $(v,w)$, then if $u=w$ go to (c),
otherwise remove $v$ from $V$ and $(u,v)$ and $(v,w)$ from
$E$ and contract $u$ and $w$ into a single node.
\abs{(b)}
If $D$ is a directed cycle of unmatched edges (so $D\subset
E-M$), remove all edges in $D$ from $E$. 
\abs{(c)} 
The two edges $(u,v)$ and $(v,u)$, one of which is matched,
form a sign-switching cycle $C$ of the reduced graph.
Repeatedly re-insert the edge pairs $(u',v')$,
$(v',w')$ removed in the contraction (a) into $C$ until $C$
is a cycle of the original graph.
Return~$C$.

\noindent
Steps (a) and (b) preserve the invariant that $G$ is an
Euler digraph and has a perfect matching.
Namely, in (a) one node and one matched and one unmatched
edge is removed from $G$, and the two contracted nodes $u$
and $w$ together have the same in- and outdegree and an
incident matched edge.
In (b), all nodes of the cycle $D$ have their in- and
outdegree reduced by~$1$.
If reduction (a) cannot be applied because every node has
at least two outgoing edges, then one of them is unmatched,
and following these edges will find a cycle~$D$ as in~(b).
So the reduction steps eventually terminate.
In each iteration in (c), the two re-inserted edges
$(u',v')$ and $(v',w')$ point in the same direction and one
of them is matched, so this preserves the property that $C$
is sign-switching.

The above algorithm is clearly polynomial.
Appendix B describes a detailed implementation with
% In V\'egh and von Stengel (2012, Appendix~B) we describe a
% detailed implementation with
near-linear running time in the
number of edges, and give an example.
Its essential features are the following.
The algorithm starts with the endpoint of a matched edge,
and follows, in forward direction, unmatched edges whenever
possible.
It there\-by generates a path of nodes connected by unmatched
edges.
If a node is found that is already on the path, then some
final part of that path forms a cycle $D$ of unmatched edges
that are all discarded as in~(b).
Then the search starts over from the beginning of the cycle
that has just been deleted.
If, in the course of this search, a node $v$ is found where
the only outgoing edge $(v,w)$ is matched, then the
contraction in (a) applies with $(u,v)$ as unmatched edge.
The matched edge $(v,w)$ is remembered as the original
matched edge incident to $w$, with $(u,v)$ as its
``partner'', for possible later re-use in (c).
The two edges are removed from the lists of incident edges
to $u$ and~$w$.  
Edges are stored in doubly-linked lists that can be moved
and deleted from in constant time.
The endpoint $w$ of the matched edge $(u,w)$ contracted in
step~(a) may be a node that has been visited on the path,
so that the reduction (b) immediately follows; if $w$
is the first node of the path, the search has to re-start.

Contracted nodes of the reduced graph are represented by 
equivalence classes of a standard {\em union-find} data
structure, which can be implemented with amortized cost
$\alpha(|V|)$ per access (Tarjan, 1975).
Contracting $u$ and $w$ in (a) is done by applying the
``union'' operation to the equivalence classes for $u$ and
$w$, and any node is represented via the ``find'' operation
applied to an original node.
The nodes in edge lists are always the original nodes,
so that each edge is visited only a constant number of
times, resulting in the running time
$O(|E|\cdot\alpha(|V|))$.

 As described in Appendix B in Figure~\ref{freconnect},
%As described by V\'egh and von Stengel (2012, Figure~10),
the cycle $C$ in (c) is obtained by recursively re-inserting
matched edges $(v',w')$ and their ``partners'' $(u',v'')$
until the nodes $v'$ and $v''$ do not just belong to the
same equivalence class (as at the time of contraction) but
are actually the same original node, $v'=v''$, of~$G$;
a similar recursion is applied to the other nodes $u'$
and~$w'$.
 Lemma~\ref{l-recon} in Appendix~B shows the correctness.
\endproof

In the remainder of this section, we consider the
complementary pivoting algorithm for perfect matchings in
Euler digraphs outlined at the end of
Section~\ref{s-labpoly}.
If $G$ is bipartite, then this algorithm terminates in
time $O(|V|)$, as noted by Merschen (2012, Lemma~4.3).
In fact, a simple extension of the pivoting method applies
to general bipartite graphs which are oriented so that the
graph has no sources or sinks (which shows that such an
orientation is not Pfaffian).

\begin{proposition}
\label{p-bip}
Consider a bipartite graph $G=(V,E)$ with an orientation so
that each node has at least one incoming and outgoing edge,
with incoming and outgoing edges stored in separate lists,
and a perfect matching $M$ of $G$.
Then a matching of opposite sign can be found in time
$O(|V|)$.
\end{proposition}

\proof
The algorithm computes a path of nodes $u_0,u_1,\ldots$ 
until that path hits itself and forms a cycle $C$, which
will be sign-switching with respect to $M$.
The edges on the path are successive matched-unmatched pairs of
edges $\{u_{2k},u_{2k+1}\}$ in~$M$ and
$\{u_{2k+1},u_{2k+2}\}$ in~$E-M$ for $k\ge0$ that point in
the same direction either as $(u_{2k},u_{2k+1})$, 
$(u_{2k+1},u_{2k+2})$ or as $(u_{2k+1},u_{2k})$, 
$(u_{2k+2},u_{2k+1})$.
Starting from any node $u_0$ and $k=0$, these are found by
following from node $u_{2k}$ its incident matched edge to
$u_{2k+1}$, where this node has an outgoing unmatched edge
to $u_{2k+2}$ in the same direction because $u_{2k+1}$ has
at least one incoming and one outgoing edge.
This repeats with $k$ incremented by one until $u_{2k+2}$ is
a previously encountered node, which is of the form $u_{2i}$
for some $0\le i<k$ because the graph is bipartite. 
Then the nodes $u_{2i},\ldots,u_{2k+2}$ define a cycle~$C$ 
which is sign-switching because it has an even number of
forward-pointing edges.
Hence, $M\triangle C$ is a matching of opposite sign to $M$.
Each node is visited at most once, so the running time is
$O(|V|)$.
\endproof

If $G$ is not bipartite, then the complementary pivoting
algorithm may have exponential running time, for any
starting node that serves as a missing label.
The construction is adapted from the exponentially long
Lemke paths of Morris (1994) for labeled {\em dual cyclic
polytopes}.
The completely labeled vertices of such polytopes correspond
to perfect matchings in Euler graphs, as noted by Casetti,
Merschen, and von Stengel (2010), in the following way.

A dual cyclic polytope is defined in any dimension $m$ with
any number $n$ of facets, $n>m$, as the ``polar polytope''
of the convex hull of $n$ points $\mu(t_j)$ on the moment
curve $\mu(t)=(t,t^2,\ldots,t^m)\T$ for $j$ in~$[n]$ (see
Ziegler, 1995).
Its vertices have been described by Gale (1963):
The $m$ facets that a vertex $x$ lies can be described by
a bit string $g=g_1g_2\cdots g_n$ in $\{0,1\}^n$ so that
$g_j=1$ if and only if $x$ is on the $j$th facet, for $j$
in~$[n]$.
Then these bit strings fulfill the {\em evenness condition}
that whenever $g$ has a substring of the form $01^k0$, then
$k$ is even.
We consider even $m$, so that these strings are preserved
under cyclical shifts.
The set $G(m,n)$  of these ``Gale strings'' encodes the
vertices of the polytope, and pivoting, and an orientation,
can be defined in a simple combinatorial way on the strings
alone.

With a labeling $l:[n]\to[m]$, the CL Gale strings therefore
come in pairs of opposite sign.
They correspond, including signs, to the {\em perfect
matchings} of the graph $G$ with node set $[m]$ and
(oriented) edges $(l(j),l(j+1))$ for $1\le j<n$ and
$(l(n),l(1))$
(Casetti, Merschen, and von Stengel, 2010;
Merschen, 2012, Theorem~3.4).
That is, the cyclic sequence $l(1),\ldots,l(n),l(1)$
defines an Euler tour of~$G$, so that $G$ is an Euler
digraph.
The graph has parallel edges and possibly loops, where the
latter can be omitted.
The 1's in a Gale string come in pairs, which correspond to
edges of~$G$.
A pivoting step from one ACL Gale string to another means
that a substring of the form $1^{2k}0$ is replaced by
$01^{2k}$, which translates to $k$ pivoting steps of skew
matchings in~$G$.
Morris (1994) gives a specific labeling for $n=2m$ where all
complementary pivoting paths, for any dropped label, are
exponentially long in~$m$.
The corresponding Euler digraph and the pivoting steps are
described in Merschen (2012, Section~4.4).

\section{Conclusions}
\label{s-conclude} 

We conclude with open questions on the computational
complexity of pivoting systems.

Consider a labeled oriented pivoting system whose components
(in particular the pivoting operation) are specified as 
polynomial-time computable functions.
Assume one CL state is given.
The problem of finding a second CL state belongs to the
complexity class PPAD (Papadimitriou, 1994).
This problem is also PPAD-complete, because finding a Nash
equilibrium of a bimatrix game is PPAD-complete (Chen and
Deng, 2006), which is a special case of an oriented pivoting
system by Proposition~\ref{p-unitv}.
However, there should be a much simpler proof of this fact
because pivoting systems are already rather general, so that
it should be possible to encode an instance of the
PPAD-complete problem ``End of the Line'' (see Daskalakis,
Goldberg, and Papadimitriou, 2009) directly into a pivoting
system.

Finding a Nash equilibrium of a bimatrix game is
PPAD-complete, and Lemke--Howson paths may be exponentially
long.
Savani and von Stengel (2006) showed this with games
defined by dual cyclic polytopes for the payoff matrices of
both players, and a simpler way to do this is to use the
Lemke paths by Morris (1994).
One motivation for the study of Casetti, Merschen, and von
Stengel (2010) was the question if finding a second
completely labeled Gale string is PPAD-complete.
This is unlikely because this problem can be solved in
polynomial time with a matching algorithm.
For the complexity class PPADS, where one looks for a second
CL state of opposite sign (Daskalakis, Goldberg, and
Papadimitriou, 2009), this problem is also solvable in
polynomial time with our algorithm of Theorem~\ref{t-euler}.

However, for room partitions of 3-oiks, already manifolds, 
finding a second room partition is likely to be more
complicated.
Is this problem PPAD-complete?
We leave these questions for further research.  

%\begin{acknowledgements}
  \subsection*{Acknowledgments}
We thank Marta Maria Casetti and Julian Merschen for
stimulating discussions during our joint research on labeled
Gale strings and perfect matchings, which led to the
questions answered in this paper.  
We also thank three anonymous referees for their helpful
comments.

%\end{acknowledgements}

%\newpage
  \section*{Appendix A: Labeling functions and Sperner Oiks}

One of the original motivations to consider room partitions
for oiks $\rooms_1,\ldots,\rooms_h$ with possibly different
dimensions  is to abstract from the original Lemke--Howson
algorithm for possibly non-square bimatrix games, which
alternates between two polytopes, represented by $\rooms_1$
and $\rooms_2$ (Edmonds, 2009).
% Similarly, the pivoting steps considered by Todd (1976)
% alternate between ``primoids'' and ``duoids'' (where a duoid
% is an oik).
Similarly, our proof of Theorem~\ref{t-sign} shows
complementary pivoting as an alternating use of the pivoting
function and the labeling function.  
Edmonds, Gaubert, and Gurvich (2010) cast the use of labels
(or ``colors'') in terms of room partitions with a special
manifold $\rooms_0$ called a {\em Sperner} oik.
If $l:V\to[m]$ is a labeling function, then the rooms of the
Sperner oik $\rooms_0$ are the {\em complements} of
completely labeled sets, that is, 
\begin{equation}
\label{R0}
\rooms_0=\{Q\subseteq V\mid |Q|=|V|-m,~l(V-Q)=[m]\}.
\end{equation} 
This is a manifold because $W$ is a wall of a room $Q$ of
$\rooms_0$ if and only if $V-W$ has $m+1$ elements of which
exactly two have the same label, so adding either element to
$W$ defines the two rooms that contain $W$.
In addition to $\rooms_0$, suppose that $\rooms$ is an
$m$-oik on $V$ and defines a pivoting system as in
Proposition~\ref{p-oikpivot}.
Then an ordered room partition $(R,Q)$ with $R\in\rooms$ and
$Q\in\rooms_0$ is just a completely labeled room $R$ of
$\rooms$.
Complementary pivoting with missing label $w$ amounts to the
``exchange algorithm'' with skew room partitions, which are
our ACL states.

Is the use of room partitions where one room comes from a
Sperner oik more natural than the concept of completely
labeled rooms?
Obviously, the definitions are nearly identical, but apart
from that we want to make two comments in favor of using
labels.

First, Edmonds, Gaubert, and Gurvich (2010) note that 
a Sperner oik $\rooms_0$ is ``polytopal'', that is, its
rooms correspond to the vertices of a simple polytope.
They leave the construction of such a polytope as an
exercise, which we give here to show the connection to
the unit-vector games in Proposition~\ref{p-unitv}.

\begin{proposition}
\label{p-R0}
Let $|V|=\{v_1,\ldots,v_n\}$ and $l:V\to[m]$ so
that $l(v_i)=i$ for $i\in[m]$.
Consider the $m\times(n-m)$ matrix
$A=[e_{l(v_{m+1})}\cdots e_{l(v_n)}]$
with $A\T=[a_1\cdots a_m]$ and
\begin{equation}
\label{P0}
P_0=\{y\in\reals^{n-m}\mid Ay\le\1,~y\ge\0\}.
\end{equation}
Then $P_0$ is a simple polytope, and $y$ is a vertex of 
$P_0$ if and only if it lies on $n-m$ facets
and the $m$ non-tight inequalities in $(\ref{P0})$ fulfill
\begin{equation}
\label{nontight}
\{i\in[m]\mid a_i\T y<1\}\cup\{l(v_{m+j})\mid y_j>0\}=[m].
\end{equation}
\end{proposition}

\proof 
For each $i$ in $[m]$ let 
\begin{equation}
\label{Li}
L(i)=\{j\in[n-m]\mid l(v_{m+j})=i\}.
\end{equation}
Then the $i$th row of $Ay\le\1$ says
$a_i\T y=\sum_{j\in L(i)}y_j\le1$.
Let $y\in P_0$.
For each $i$, if $a_i\T y=\sum_{j\in L(i)}y_j=1$, then
$y_j>0$ for at least one $j$ in $L(i)$, so
$i\in\{l(v_{m+j})\mid y_j>0\}$, which shows
(\ref{nontight}).

The non-empty sets $L(i)$ form a partition of $[n-m]$,
and if $L(i)$ is empty then $a_i=\0$ and the inequality
$a_i\T y\le1$ is redundant.
Therefore the inequalities (\ref{P0}) can be re-written as
\begin{equation}
\label{simplex}
\sum_{j\in L(i)}y_j\le 1,
\quad
y_j\ge0 ~~(j\in L(i)).
\end{equation}
For each $i$ in $[m]$, (\ref{simplex}) defines a simplex
whose vertices are the unit vectors and $\0$ in
$\reals^{|L(i)|}$ (if $L(i)$ is empty, this is the one-point
simplex $\{()\}$).
Hence, $P_0$ is the product of these simplices and therefore
a simple polytope,
so any vertex $y$ of $P_0$ is on exactly $n-m$ facets.
\endproof

Proposition~\ref{p-R0} can be applied to any Sperner oik
$\rooms_0$ of dimension $n-m$ obtained from $l:V\to[m]$
which has at least one room, taken to be
$\{v_{m+1},\ldots,v_n\}$ by numbering $V$ suitably.
The $n$ inequalities in (\ref{P0}) have labels
$1,\ldots,m,l(v_{m+1}),\ldots,l(v_m)$; they define facets
of $P_0$ except for redundant inequalities $a_i\T y\le1$
where $a_i=\0$.
Then the $n-m$ tight inequalities for each vertex $y$ of
$P_0$ define a room of $\rooms_0$ because the labels
for the $m$ non-tight inequalities for $y$ are the set $[m]$
according to (\ref{nontight}), in agreement with~(\ref{R0}).
% the definition of the Sperner oik $\rooms_0$.

Suppose $\rooms$ is an $m$-oik given by the vertices of the
polytope $P$ in (\ref{Pge0}), with labels
$1,\ldots,m,l(m+1),\ldots,l(n)$ for its $n$ inequalities
(the same labels as for $P_0$).
Then an ordered room partition $R,Q$ with $R\in\rooms$ and
$Q\in\rooms_0$ is a completely labeled room $R$, or vertex
$x$ of $P$, with $Q$ corresponding to a vertex $y$ of $P_0$.
Except for the vertex pair $(\0,\0)$, this is a Nash
equilibrium $(x,y)$ of the unit-vector game $(A,C\T)$ in
Proposition~\ref{p-unitv}.
In that game, there is no reference to labels, which are
encoded in the payoff matrix $A$ that defines $P_0$, just as
the labels are encoded in the rooms of $\rooms_0$.
% and ``completely labeled'' as ``room partition''.
Like unit vector games, Sperner oiks may offer a useful
perspective, but we do not think it is deep;
moreover, they only have a simple structure as products of
simplices described in (\ref{simplex}).

\begin{figure}[hbt] 
\strut\hfill
\epsfxsize7cm
\epsfbox{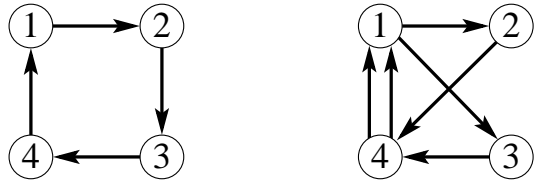} 
\hfill
\strut 
\caption{Two oriented Euler graphs which show that the
parity of the permutation of all nodes matters.
} 
\label{4nodes}
\end{figure} 

Secondly, Sperner oiks are oriented, and the labels used in
the proof of Theorem \ref{t-ordpart} and~\ref{t-part} are
simply the nodes of~$V$.
Perhaps using a Sperner oik, rather than labels, may
avoid referring to the parity of $l(r(s))$ for a room
partition $s$ as in (\ref{CLsign}) when defining the sign
of~$s$?
The following example shows that already when $s$ is a room
partition for a 2-oik, one has to refer to the parity of
$l(r(s))$ in some way.
Figure~\ref{4nodes} shows two cases of 2-oiks $\rooms'$ over
$V=\{1,2,3,4\}$ with an orientation.
The left oik has the two room partitions $\{12,
34\}$ and $\{14, 23\}$, where $\sigma_1(12)=\sigma_2(34)=1$
and $\sigma_1(14)=-1$, $\sigma_2(23)=1$.
% (only $14$ is oriented against the edge $41$).
According to (\ref{sumorient}), this implies
$\sigma(12,34)=1$ and $\sigma(14,23)=-1$, so the two room
partitions have opposite orientation (it suffices to
consider unordered room partitions because $d$ is even, as
noted in Theorem~\ref{t-part}).

Similarly, the right oik in Figure~\ref{4nodes} has the two
room partitions $\{12, 34\}$ and $\{13, 24\}$, where
$\sigma_1(12)=\sigma_2(34)=1$ and
$\sigma_1(13)=\sigma_2(24)=1$, so all orientations are
positive and $\sigma(12,34)=1$ and $\sigma(13,24)=1$, so
these two room partitions have equal orientation.  The
difference is that the room partition $s=\{14,23\}$ defines an
even permutation $l(r(s))=(1,4,2,3)$ of $V$, whereas
$\{13,24\}$ defines the odd permutation $(1,3,2,4)$.
So the sign of a room partition has to refer to the order in
which the labels appear.

We think that labeled pivoting systems are a general
and useful way of representing path-following and parity
arguments, certainly for complementary pivoting and room
partitions in oiks.

  \section*{Appendix B: Implementation Details of Finding a
Sign-Switching Cycle in an Euler Graph}

Theorem~\ref{t-euler} states that an oppositely signed
matching in a graph with an Eulerian orientation can be
found in near-linear time in the number of edges.
In this appendix, we describe the details of the
implementation of the algorithm outlined in the proof of
Theorem~\ref{t-euler}.

When $e$ is an edge from $u$ to $v$, then we call $u$ the
{\em tail} and $v$ the {\em head} of~$e$, and both $u$ and
$v$ are called {\em endpoints} of~$e$.

The algorithm applies reductions (a) and (b) to the graph
until it has a trivial sign-switching cycle which is expanded
as in (c) to form a sign-switching cycle of the original
graph.
The algorithm starts with a node that is the head of a
matched edge, and follows, in forward direction, unmatched
edges whenever possible.
% It thereby generates a path of nodes $u_1u_2\ldots u_k$
% connected by unmatched edges
% $u_iu_{i+1}$ for $1\le i<k$.
% As soon as a node $u_k$ is found that is already on the
% path, that is, $u_k=u_j$ for some $j<k$, then the edges 
% $u_iu_{i+1}$ for $j\le i<k$ form a cycle of unmatched edges
% that are all discarded as in (b).
% Then the search starts over from~$u_j$.
It thereby generates a path of nodes connected by unmatched
edges.
If a node is found that is already on the path, then some
final part of that path forms a cycle $D$ of unmatched edges
that are all discarded as in~(b).
Then the search starts over from the beginning of the cycle
that has just been deleted.

%\abs{(2)} 
If, in the course of this search, a node $v$ is found with the
only outgoing edge being matched, the contraction in (a) is
performed as follows.
Suppose the three nodes in question are $u,v,w$ with
unmatched edge $e$ from $u$ to~$v$ and matched edge $m$ from
$v$ to~$w$, and no other edge incident to $v$.
We take the edges $e$ and $m$ and node $v$ out of the graph
and contract the nodes $u$ and $w$ into a single node (with
the method $\shrink(e,m)$ discussed below), which creates a
reduced version of the graph.
Throughout the computation, the current reduced graph is
represented by a {\em partition} of the nodes with a
standard {\em union-find} data structure (Tarjan, 1975).
We denote by $[x]$ the partition class that contains
node~$x$, which has as its {\em representative} a special
node called $\find(x)$, where $\find$ is one of the standard
union-find methods; we usually denote a representative node
with a capital letter.  
That is, any two nodes $x$ and $y$ are equivalent (in the
same equivalence class) if and only if $\find(x)=\find(y)$.
In the reduced graph, {\em every edge} is only incident to
the representative $\find(x)$ of a partition class,
%%%%%%%%%% rest of sentence possibly not clear
and the information for nodes that are not representatives
is irrelevant.
Initially, all partition classes are singletons $\{x\}$,
which is achieved by calling the $\makeset(x)$ method.
The method $\unite(x,y)$ for nodes $x,y$ merges $[x]$ and
$[y]$ into a single set.

\begin{figure}[hbt] 
\strut\hfill
\fbox{\hsize=27.3em\vbox{%
\vspace{2mm}
\noindent
\strut~~~$\makeset(x)$:
\\[1mm]
\4 $x.\parent\SET x$
\\
\4 $x.\rank\SET 0$
\\[2mm]
\noindent
\strut~~~$\unite(x,y)$:
\\[1mm]
\4 $X,Y\SET \find(x),\find(y)$
\\
\4 \IF{$X.\rank > Y.\rank$}
\\
\4\3 $Y.\parent\SET X$
\\
\4\3 $\return X,Y$
\\
\4 \ELSE
\\
\4\3 $X.\parent\SET Y$
\\
\4\3 \IF{$X.\rank = Y.\rank$}
\\
\4\3\3 $Y.\rank \SET Y.\rank+1$
\\
\4\3 $\return Y,X$
\\[2mm]
\noindent
\strut~~~$\find(x)$:
\\[1mm]
\4 \IF{$x\ne x.\parent$}
\\
\4 \3 $x.\parent\SET \find(x.\parent)$
\\
\4 \return $x.\parent$
\vspace{2mm} 
}}
\hfill
\strut 
\caption{The union-find methods \makeset, \unite, and
\find{} with rank heuristic and path compression.
Here, $\unite(x,y)$ returns $X,Y$ so that $X$ is the new
representative of $[x]\cup [y]$, and $Y$ is the old
representative of either $[x]$ or $[y]$ which is no longer
used.} 
\label{funionfind}
\end{figure} 

Figure~\ref{funionfind} shows an implementation of these
methods as in Cormen et al.\ (2001, Section 21.3).  
(In this pseudo-code, an assignment such as $X,Y\SET x,y$
assigns $x$ to $X$ and $y$ to $Y$, so for example $x,y\SET
y,x$ would exchange the current values of $x$ and~$y$.)
Each partition class is a tree with $x.\parent$ pointing to
the tree predecessor of node~$x$, which is equal to~$x$ if
$x$ is the root.
For this root, $x.\rank$ stores an upper bound on the
height of the tree.
The $\unite$ method returns the pair $X,Y$ of former
representatives of the two partition classes, where $X$ is
the new representative of the merged partition class and $Y$
is the representative no longer in use, which we need in
order to move edge lists in the graph.
With the ``rank heuristic'' used in the $\unite$ operation
and the ``path compression'' of the recursive $\find$
method, the trees representing the partitions are extremely
flat, with an amortized cost for the \find\ method given by
the inverse Ackermann function that is constant for all
conceivable purposes (see Tarjan, 1975, and Cormen et al.,
2001, Section 21.3).

Every node $x$ of the graph has its incident edges stored in
an {\em adjacency list}, which for convenience is given by
separate lists $x.\outlist$ and $x.\inlist$ for unmatched
outgoing and incoming edges, respectively, and the unique
matched edge $x.\matched$ which is either incoming or
outgoing.  
Every edge $e$ is stored in a single object that contains
the following links to edges:
$e.\nextout$,
$e.\prevout$,
$e.\nextin$,
$e.\previn$,
which link to the respective next and previous element in the
doubly-linked $\outlist$ and $\inlist$ where $e$ appears.
In addition, $e$ contains the links to two nodes 
$e.\tail$ and $e.\head$, which never change, 
so that $e$ is always an edge from $e.\tail$ to $e.\head$ in
the {\em original} graph.
In the current reduced graph at any stage of the
computation, $e$ is an edge from $\find(e.\tail)$ to
$\find(e.\head)$, so these fields of $e$ are not updated
when $e$ is moved to another node in an edgelist; this
allows to move all incident edges from one node to another
in constant time.

\begin{figure}[hbt] 
\strut\hfill
\fbox{\hsize=27.3em\vbox{%
\vspace{2mm}
\noindent
\strut~~~$\shrink(e,m)$:
\\[1mm]
\5 $U,W\SET \find(e.\tail) ,\find(m.\head)$
\\
\5 remove $e$ from $U.\outlist$
\\
%\6 [ $V\SET \find(e.\head)$ ]
\6 $V\SET \find(e.\head)$
\\
\6 $\sleepcounter\SET\sleepcounter+1$
\\
\6 $V.\sleeptime\SET\sleepcounter$
\\
\5 $m.\partner\SET e$
\\
\5 $X,Y\SET \unite(U,W)$
\\
\5 append $Y.\outlist$ to $X.\outlist$
\\
\5 append $Y.\inlist$ to $X.\inlist$
\\
\5 $X.\matched\SET U.\matched$
\vspace{2mm} 
}}
\hfill
\strut 
\caption{The $\shrink$ operation that removes the
unmatched edge $e$ from $U$ to $V$ and matched edge $m$ from
$V$ to $W$ from the current graph and merges the edgelists
of $U$ and~$W$.
The code in the starred lines 3--5 is only needed to
reason about the method and can be omitted.
} 
\label{fshrink}
\end{figure} 

\begin{figure}[hbt] 
\strut\hfill
\epsfysize11em
\epsfbox{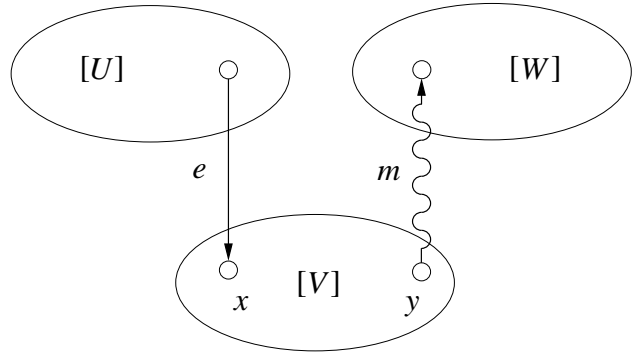} 
\hfill
\strut 
\caption{The equivalence classes $[U]$, $[V]$, $[W]$ and
edges $e$ and $m$ in the $\shrink$ operation.
A wiggly line denotes a matched edge.
} 
\label{picshrink}
\end{figure}

Figure~\ref{fshrink} gives pseudo-code for the contraction
(a) described above.
The three nodes $U,V,W$ are the representatives of their
partition classes, and only for these nodes the lists of
outgoing and incoming edges and their matched edge are
relevant.
The unmatched edge $e$ appears in $U.\outlist$ and has head
$V$, so that $U=\find(e.\tail)$ and
$V=\find(e.\head)=\find(m.\tail)$, even though
it may be possible that $e.\head\ne m.\tail$ like for $x,y$
in Figure~\ref{picshrink}.
The matched edge $m$ from $V$ to $W$ is obtained as
$V.\matched$ (and equals $W.\matched$), because $V.\outlist$
is empty so $V$ has no outgoing unmatched edge (but has to
have an outgoing edge due to the Eulerian orientation).

After the $\shrink$ operation, the reduced graph no longer
contains the edges $e$ and $m$ and the node $V$.
(However, these are preserved for later re-insertion, helped
by the field $m.\partner$ assigned to $e$ in line~6 of
$\shrink$, discussed below along with lines 3--5.)
The edge $e$ is removed from the list of outgoing edges of
$U$ in line~2.
The equivalence classes for $U$ and $W$ are united in line~7
where either $U$ or $W$ becomes the new representative,
stored in~$X$.
The lists of outgoing and incoming edges of the
representative $Y$ that is no longer in use are appended to
those of $X$ in lines 8 and~9.
A node can only lose but never gain the status of being a
representative, so there is no need to delete the edgelists
of~$Y$.
If the new representative $X$ is $W$, its current matched
edge $m$ has to be replaced by the matched edge $U.\matched$
as in line~10 (which has no effect if $X=U$).
The Euler property of the reduced graph is preserved because
the outdegree of $X$ is the sum of the outdegrees of $U$ and
$W$ minus one, and so is the indegree (the missing edges are
$e$ and $m$).

\begin{figure}[hbt] 
\strut\hfill
\epsfxsize31em
\epsfbox{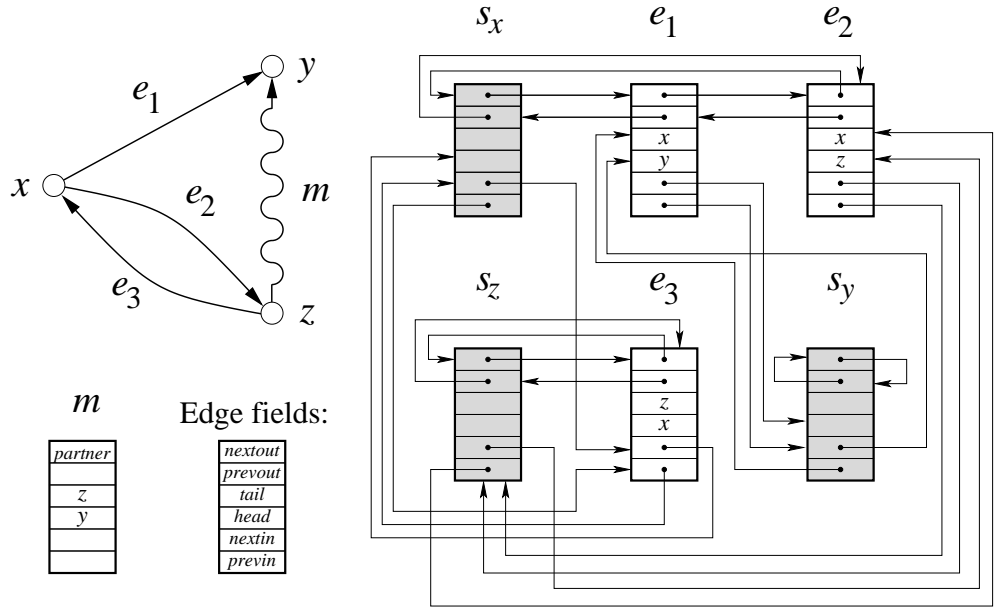} 
\hfill
\strut
\caption{Example of a graph with the out- and inlists for
the nodes $x,y,z$ accessed by sentinels (dummy edges)
$s_x,s_y,s_z$ shown in gray. 
They use the same fields $\nextout$, $\prevout$, $\nextin$,
$\previn$ as the unmatched edges $e_1,e_2,e_3$ except for
$\tail$ and $\head$ which are ignored.
The matched edge $m$ is stored directly, linked to by
$y.\matched$ and $z.\matched$, and not in a list.
The $m.\nextout$ field can be used to link to $m.\partner$.
} 
\label{piclist}
\end{figure} 

The list operations in lines 2, 8, 9 of $\shrink$ can be
performed in constant time.
For that purpose, it is useful to store the lists of
outgoing and incoming unmatched edges of a node $u$ as
doubly-linked circular lists that start with a ``sentinel''
(dummy edge), denoted by $s_u$ (see Cormen et al., 2001,
Section~10.2).  
Figure~\ref{piclist} gives an example of small graph (which
is neither Eulerian nor has a perfect matching).
The three unmatched edges are $e_1,e_2,e_3$ and the matched
edge is $m$.
The outlist of $x$ contains $e_1,e_2$,
the outlist of $y$ is empty, 
and the outlist of $z$ contains $e_3$.
The inlist of each node contains exactly one edge.
The first and last element of the outlist of a node
$u$ is pointed to by $s_u.\nextout$ and $s_u.\prevout$,
which are both $s_u$ itself when the list is empty, as 
for $u=y$ in the example.
The inlist is similarly accessed via $s_u.\nextin$ and
$s_u.\previn$.
Each append operation in line 8 or 9 of $\shrink$ is then
performed by changing four pointers.
The remove operation in line 2 can, in fact, be done
directly from $e$, again by changing four pointers, here
of the next and previous edge in the list (which may be a
sentinel).
Due to the sentinels, $e$ does not need the information of
which node it is currently attached to, so line~2 should
be written (a bit more obscurely) as ``remove $e$ from its
outlist'' (that is, the outlist it is currently contained
in), without reference to~$U$.

\begin{figure}[hbtp] 
\strut\hfill
\fbox{\hsize=27.3em\vbox{%
\vspace{2mm}
\noindent
\strut~~~$\fosm$:
\\[1mm]
\5 \FOR{} all nodes $u$
\\
\5\3 $\makeset(u)$
\\
\5\3 $u.\origmatched\SET u.\matched$
\\
\5\3 $u.\visited\SET 0$
\\
\6\3 $u.\sleeptime\SET 0$ 
\\
\6 $\sleepcounter\SET 0$ 
\\[1mm]
\MARK{A} $m\SET$ any matched edge of current graph
\\
\5 $V\SET \find(m.\head)$
\\
\5 $\vc\SET 1$
\\[1mm]
\MARK{B} $\vn[\vc]\SET V$
\\
\5 $V.\visited\SET\vc$
\\
\5 \IF{$V.\outlist$ is not empty}
\\
\5\3 $e\SET$ first edge in $V.\outlist$
\\
\5\3 $W\SET \find(e.\head)$
\\
\5\3 $\ve[\vc]\SET e$
\\
\5\3 $\vc\SET \vc+1$
\\
\5\3 $\checkvisited(W)$
\\
\5\3 $V\SET W$
\\
\5\3 $\GOTO B$
\\
\5\ELSE
\\
\5\3 $m\SET V.\matched$
\\
\5\3 $W\SET \find(m.\head)$
\\
\5\3 $\vc\SET \vc-1$
\\
\5\3 $U,e \SET \vn[\vc], \ve[\vc]$
% \5\3 $U\SET \vn[\vc]$
% \\
% \5\3 $e\SET \ve[\vc]$
\\
\5\3 \IF{$W=U$} 
\\
\5\3\3 \return{$\expandcycle(e,m)$}
\\
\5\3 $\shrink(e,m)$

\5\3 $\checkvisited(W)$
\\
\5\3 \IF{$\vc>1$}
\\
\5\3\3 $V\SET \find(W)$
\\
\5\3\3 $\GOTO B$
\\
\5\3\ELSE
\\
\5\3\3 $\GOTO A$

\vspace{2mm} 
}}
\hfill
\strut 
\caption{The main method $\fosm$ for an Euler graph
with a given perfect matching.
} 
\label{fmain}
\end{figure} 

Figure \ref{fmain} shows the whole algorithm that finds a
perfect matching of opposite sign via a sign-switching
cycle.
Initialization takes place in lines 1--6, which will be
explained when the respective fields and variables are used.

The main computation starts at step {\bf A}.
The first node $V$ is the head of a matched edge.
This assures that, due to the Euler property, this node has
at least one outgoing unmatched edge that may be the first
edge $e$ of an edge pair $e,m$ that is contracted with the
$\shrink$ method.
Starting from step {\bf B}, a path of unmatched edges is
grown with its nodes stored in $\vn[1],\ldots,\vn[\vc]$
where $\vc$ counts the number of visited nodes, and
edges stored in $\ve[1],\ldots,\ve[\vc-1]$,
where $\ve[i]$ is the %unmatched
edge from $\vn[i]$ to $\vn[i+1]$ for $1\le i<\vc$.
A node $u$ is recognized as visited on that path when
$u.\visited$ is positive, which is the index $i$ so that
$u=\vn[i]$.
This field is initialized in line~4 as initially zero
(unvisited).

Line 10 tests if $V$ has a non-empty list of outgoing
unmatched edges, which is true when $\vc=1$.
The next node, following the first edge $e$ of that list,
is~$W$.
Line 15 checks with the method $\checkvisited$, shown in
Figure~\ref{fcheck}, if $W$ has been visited before.
If that is the case, then all edges in the corresponding
cycle are completely removed from the graph and the nodes
are marked as unvisited (lines 3 and 4 of $\checkvisited$ in
Figure~\ref{fcheck}), and $\vc$ is reset to the beginning of
that cycle.
In any case, $W$ is the next node of the path, and the loop
repeats at step~{\bf B} via line~17.

\begin{figure}[hbt] 
\strut\hfill
\fbox{\hsize=27.3em\vbox{%
\vspace{2mm}
\noindent
\strut~~~$\checkvisited(W)$:
\\[1mm]
\5 \IF{$W.\visited > 0$}
\\
\5\3 \FOR\ $i\SET W.\visited,\ldots,\vc-1$
\\
\5\3\3 remove $\ve[i]$ from its $\outlist$ and $\inlist$
\\
\5\3\3 $\vn[i].\visited\SET 0$
\\
\5\3 $\vc\SET W.\visited$
\vspace{2mm} 
}}
\hfill
\strut 
\caption{The $\checkvisited$ method that
checks if node $W$ has already been visited, and if yes
deletes the encountered cycle of unmatched edges and
updates~$\vc$.
} 
\label{fcheck}
\end{figure} 

\begin{figure}[t] 
\strut\hfill
\epsfysize11em
\epsfbox{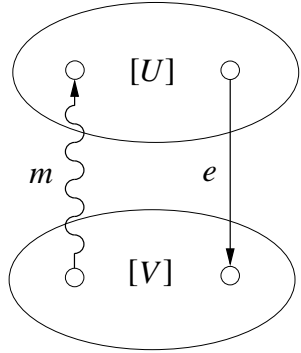} 
\hfill
\strut 
\caption{The equivalence classes $[U]$, $[V]$ when a
sign-switching cycle has been found.
} 
\label{picsscycle}
\end{figure}

Lines 18--31 deal with the case that $V$ has no outgoing
unmatched edge, which can only hold if $\vc>1$.
Then the matched edge $m$ incident to $V$ is necessarily
outgoing due to the Euler property and because $V$ has an
incoming edge $e$ from $U$ to $V$, which is found in line~22.
This edge is normally removed in the $\shrink$ operation and
then no longer part of the path, which is why $\vc$ is
decremented in line~21 (node $V$ will no longer be part of
the graph and can keep its $\visited$ field).
However, a sign-switching cycle is found if $W=U$ (see
Figure~\ref{picsscycle}), which is tested in line~23 and
dealt with in the $\expandcycle$ method called in line~24,
which terminates the algorithm and will be explained below.

If $W\ne U$, then $\shrink(e,m)$ is called in line~25.
Afterwards, node $W$ is still the old representative of the
head node of~$m$, as it was used in finding the path of
unmatched edges. 
Node $W$ may be part of that path, as tested (and the
possible cycle removed) in line~26.
If $W$ has been visited, then $W.\visited<\vc$ because
$W\ne U$, so at least one edge is removed.
If $\vc>1$ (which holds in particular if $W$ has not been
visited), then the path is now grown from $\find(W)$ in
line~28, where the $\find$ operation is needed to update
$\vn[\vc]$ in step~{\bf B} because the old representative
$U$ may have been changed to $W$ after the $\unite$
operation in line~7 of $\shrink$ in Figure~\ref{fshrink}.

The case $W.\visited=1$ needs special treatment, which
results in $\vc=1$ and happens when $m$ is same as the
initial matched edge $m$ in step~{\bf A}.
In that case, $m$ is removed via $\shrink$, and $\find(W)$
may now be the tail rather than head of a matched edge,
and then may no longer have an unmatched outgoing edge,
which is necessary for lines 21--22 to work.
For that reason, the loop goes back to~{\bf A} rather
than {\bf B}, as in line~31; note that $W.\visited$ has been
set to zero in line~4 of \checkvisited\ with $i=\vc=1$.
In step~{\bf A}, a new matched edge that has not yet been
removed in a call to \shrink\ can be found in constant time
by storing all matched edges in a doubly-linked list (for
example, using the fields $\nextin$ and $\previn$ that so
far are unused for matched edges, see Figure~\ref{piclist});
a matched edge $m$ should be deleted from that list after
line~6 of \shrink\ in Figure~\ref{fshrink}, for example.

\begin{figure}[hbt] 
\strut\hfill
\fbox{\hsize=27.3em\vbox{%
\vspace{2mm}
\noindent
\strut~~~$\expandcycle(e,m)$:
\\[1mm]
\5 $C\SET\{(e,m)\}$
\\
\5 $\reconnect(e.\head,m.\tail,C)$
\\
\5 $\reconnect(e.\tail,m.\head,C)$
\\
\5 \FOR\ all $(e,m)\in C$
\\
\5\3 make $e$ a matched edge and $m$ an unmatched edge
\\
\5 \return graph with this new matching 
\\[3mm]
\strut~~~$\reconnect(x,y,C)$:
\\[1mm]
\5 \IF{$x\ne y$}
\\
\5\3 $m\SET x.\origmatched$
\\
\5\3 $e\SET m.\partner$
\\
%\5\3 add $(e,m)$ to $C$
\5\3 $C\SET C\cup\{(e,m)\}$
\\
\5\3 $\reconnect(e.\head,m.\tail,C)$
\\
\5\3 $\reconnect(e.\tail,y,C)$
\vspace{2mm} 
}} 
\hfill
\strut 
\caption{The \expandcycle\ and the recursive
\reconnect\ method that create the sign-switching cycle
and with it the oppositely signed matching.
} 
\label{freconnect}
\end{figure} 

We now discuss how to re-insert the contracted edges into
the graph once a sign-switching cycle has been found,
which is done in the \expandcycle\ method in
Figure~\ref{freconnect}.
The method itself is straightforward.
Recall that edges of the current graph are stored with the
representatives of equivalence classes, where an unmatched
edge is accessed in line~11 and a matched edge in line~19 of
the main method \fosm\ in Figure~\ref{fmain}.
The sign-switching cycle will be reconstructed using the
original endpoints of the edges.
For each node $u$, the original matched edge incident to $u$
is stored in $u.\origmatched$ (see line~3 of the main
method), because $u.\matched$ may be modified (in line~10 of
the $\shrink$ method).

In order to explain the \expandcycle\ method, we record the
time at which the \shrink\ operation has been applied to a
node $V$.
This is done in lines 3--5 in Figure~\ref{fshrink} using
using the field $V.\sleeptime$ and the global variable
$\sleepcounter$, which are initialized in lines 5--6 of
Figure~\ref{fmain}.
These lines have a ``star'' to indicate that they do not
affect the algorithm, and can therefore be omitted. 
We use them to reason about the correctness of the
\expandcycle\ method.

The contraction $\shrink(e,m)$ affects three equivalence
classes $[U]$, $[V]$, $[W]$ with representatives $U,V,W$ as
shown in Figure~\ref{picshrink}.
All nodes in $[V]$ become inaccessible afterwards, but the
equivalence class still exists (and is in fact still
represented in the union tree by those nodes $v$ so that
$V=\find(v)$, although the union-find data structure will no
longer be used for these nodes).
We say that all nodes in $[V]$ become {\em asleep} at the
time recorded in the positive integer $V.\sleeptime$.
Any node $u$ so that $\find(u).\sleeptime=0$ is called
{\em awake}.
\einr1.8em

\begin{lemma}
\label{l-awake}
During the main method \fosm, the following condition holds
after any statement from step {\bf A} onwards. 
Let $[U]$ be an equivalence class of nodes with
representative $U$ and let $m'=U.\matched$.
Then there is exactly one node $u$ in $[U]$ and another
node $z$ not in $[U]$ so that:
\abs{(i)}
If $U$ is awake, then $z$ is awake and
$\{u,z\}=\{m'.\head,m'.\tail\}$.
\abs{(ii)}
If $U$ is asleep, then $z$ is awake or asleep with later
sleeptime than $U$, and $u=m'.\tail$, $z=m'.\head$.

\noindent
In either case:
\abs{(iii)}
For every node $y$ in $[U]-\{u\}$, let $m=y.\origmatched$.
Then $y=m.\head$, the node $m.\tail$ is asleep (with earlier
sleeptime than $U$ if $U$ is asleep), there is an edge $e$
so that $e=m.\partner$, the nodes $m.\tail$ and $e.\head$
are equivalent and with $x=e.tail$ we have
$x\in [U]-\{y\}$.
\end{lemma}

\begin{figure}[hbt] 
\strut\hfill
\epsfysize11em
\epsfbox{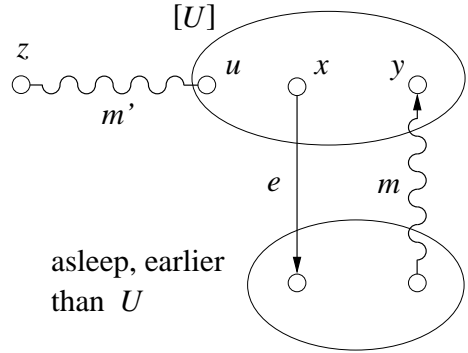} 
\hfill
\strut 
\caption{Illustration of Lemma~\ref{l-awake}.
The matched edge $m'$ with endpoints $u$ and $z$ may have
either orientation if $u$ is awake, otherwise $u=m'.\tail$
as stated in (ii).
} 
\label{fawake}
\end{figure} 

\proof
We prove this by induction over the number of calls to the
\shrink\ method, which are the only times when the
equivalence classes change.
Initially, all equivalence classes are singletons and all
nodes are awake.
Then $[U]=\{U\}$, only (i) applies, where $m'$ is the
matched edge incident to $u$ which is either the tail or
head of $m'$ and $z$ is the other endpoint of $m'$, and
(iii) holds trivially.

Figure~\ref{fawake} shows the general case of the lemma.
Consider the three equivalence classes $[U]$, $[V]$, $[W]$
with representatives $U,V,W$ as shown in
Figure~\ref{picshrink} in the notation used for the \shrink\
method.
Before $\shrink(e,m)$ is called, the lemma applies by
inductive assumption to each of the three classes $[U]$,
$[V]$, $[W]$ in place of $[U]$.
The unique matched edge $m'$ that goes outside the
equivalence class to an awake node is $m$ for $[V]$ and
$[W]$, and for $[U]$ it is some other matched edge $m'$
(not shown in Figure~\ref{picshrink}) which will be that
edge after $[U]$ and $[W]$ have been united.
There is no edge other than $e$ or $m$ from a node in $[V]$
to an awake node outside $[V]$ because only in this case 
(when $V$ has in- and outdegree one in the reduced graph)
the \shrink\ method is called.
Every node in $[U]$, $[V]$, or $[W]$ is the endpoint of a
matched edge in the original graph, and other than the
endpoints of $m$ and $m'$ they are all equal to the head of
such a matched edge, with its tail node asleep, by inductive
assumption (iii).

After the \shrink\ operation, $[U]$ and $[W]$ become a
single equivalence class, and all nodes in $[V]$ becomes
asleep.
The only node $y$ in the new class $[U]\cup [W]$ for which
(iii) does not hold by inductive assumption is $m.\head$,
but then $e$ takes exactly the described role as
$m.\partner$.
In particular, $x=e.\tail\ne y$, because $x\in [U]$ and
$y\in[W]$ and $[U]\ne[W]$.
In addition, $[V]$ changes its status from awake to asleep,
and all nodes in $[V]-\{m.\tail\}$ are heads of matched edges
that connect to equivalence classes that went asleep before~$V$
as claimed in (iii) by the inductive hypothesis.  
This completes the induction.
\endproof

The previous lemma implies that any two endpoints of a
matched edge belong to different equivalence classes.
A key observation in (iii) is that for any $y$ in an
equivalence class $[U]$ that is not the endpoint $u$ of the
``awake'' matched edge $m'$ there is another node $x$ {\em
different} from $y$ in that class (which may be $u$) given
by
%$$x=e.\tail$.
$x=y.\origmatched.\partner.\tail$.

\begin{figure}[hbt] 
\strut\hfill
\epsfysize14.4em
\epsfbox{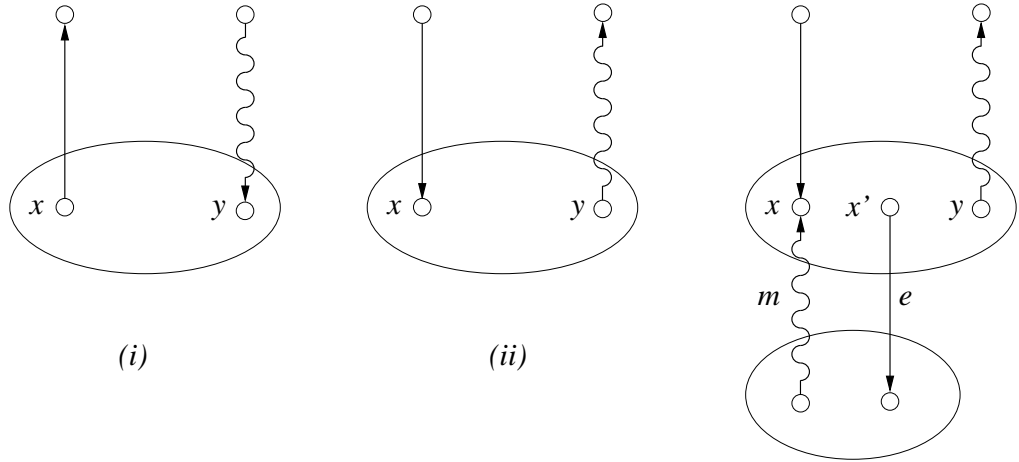} 
\hfill
\strut 
\caption{Illustration of Lemma~\ref{l-recon} about the
\reconnect\ method.
} 
%\begin{figure}[hbt] 
\label{frecon}
\end{figure}

\begin{lemma}
\label{l-recon}
Consider nodes $x,y$ and a set $C$ with the following
properties:
$x$ and $y$ are equivalent, and $x$ is the endpoint of an
unmatched edge and $y$ is the endpoint of an oppositely
oriented matched edge taken from the pairs of
unmatched-matched edge pairs in~$C$, as in (i) or (ii) in
Figure~\ref{frecon}.
Then after $\reconnect(x,y,C)$, the new edges in $C$ form a
path of alternating matched-unmatched edges from $x$ to $y$
with the same number of matched and unmatched
forward-pointing edges.
\end{lemma}

\proof
If $x=y$, then the claim holds trivially with a path of zero
length because no edge pairs are added to~$C$.
Otherwise, we apply Lemma~\ref{l-awake} to the equivalence class that
contains $x$ and $y$, as shown in the right picture in
Figure~\ref{frecon}, where $x$ takes the role of $y$ in
Lemma~\ref{l-awake}(iii).
Hence, node $x$ is the head of some matched edge $m$
with partner $e$, and $x'=e.\tail\ne x$.
Then in the method \reconnect\ (line 10 in
Figure~\ref{freconnect}), $(e,m)$ is added to $C$.
We then apply the claim recursively to $e.\head,m.\tail$
instead of $x,y$, and then to $x',y$ instead of $x,y$, where
the assumptions apply, exactly as in lines 11 and 12 of
\reconnect.
So there are alternating paths as described from $e.\head$ to
$m.\tail$ and from $x'$ to~$y$.
The resulting path from $x$ to $y$ composed of these paths
and the edges $m$ and $e$ has the same number of 
forward-pointing matched and unmatched edges, because 
$m$ and $e$ point in the same direction (in this case,
backwards) along the path.
\endproof

In the \expandcycle\ method, lines 2 and 3 in
Figure~\ref{freconnect} call $\reconnect(x,y,C)$ for the
endpoints $x,y$ of the unmatched-edge pair shown in
Figure~\ref{picsscycle} that results when \expandcycle\ is
called from the main method (line 24 of Figure~\ref{fmain}),
first for $x,y$ in $[V]$ and then for $x,y$ in $[U]$ in
Figure~\ref{picsscycle}.
In both cases, Lemma~\ref{l-recon} applies, and the
paths together with the first edge pair $(e,m)$ form a
sign-switching cycle.

Finally, exchanging the matched and unmatched edges as in
line~5 of \expandcycle\ can be done as described,
irrespective of the order of the edges in the cycle, 
just using the pairs $(e,m)$ in $C$ (which are oriented in
the same direction along the cycle), which suffices to
obtain a matching of opposite sign. 

This concludes the detailed description of the algorithm.
It has near-linear running time in the number of edges,
because each unmatched edge is visited at most once and
either discarded or contracted in the course of the
algorithm.  

\begin{figure}[hbt] 
\strut\hfill
\epsfysize14em
\epsfbox{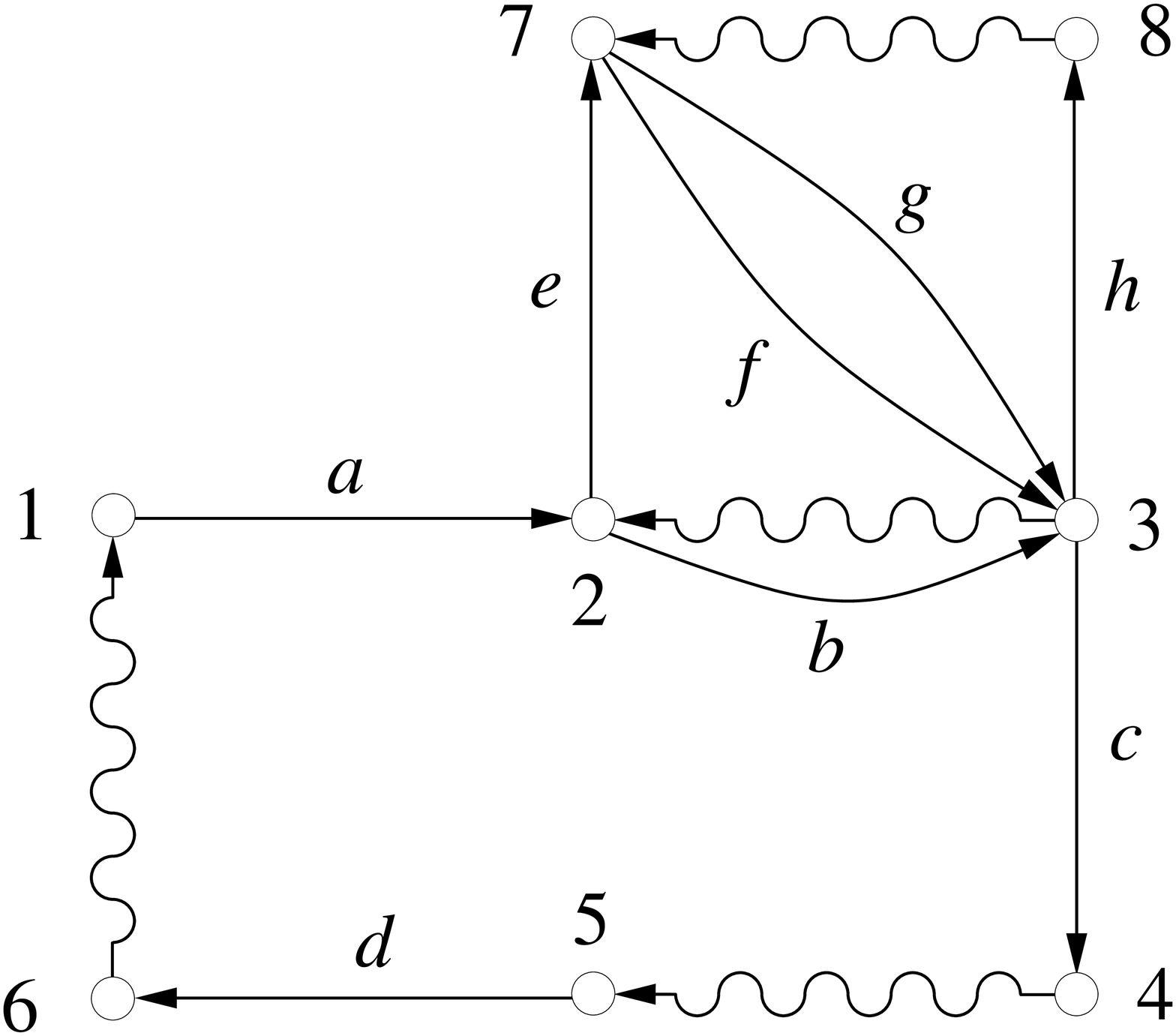} 
\hfill
\strut 
\caption{Example to illustrate the algorithm.
Unmatched edges are marked $a$ to~$h$, matched edges are
identified by their endpoints.
The first matched edge is $61$.}
\label{fex}
\end{figure} 

\begin{figure}[hbt] 
\strut\hfill
\epsfysize14em
\epsfbox{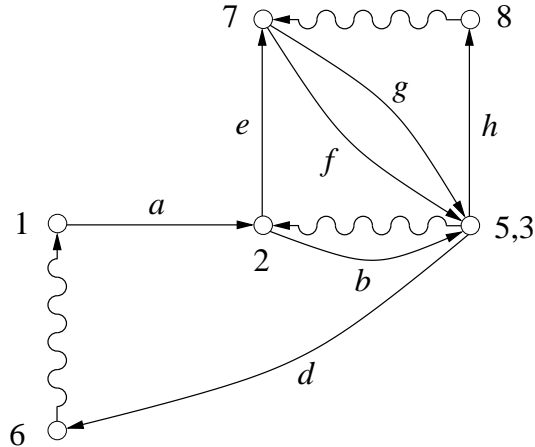} 
\hfill
\strut 
\caption{The reduced graph after the first contraction with
$\shrink(c,45)$.
The equivalence class with nodes $5,3$ is written with the
representative listed first.
}
\label{fex1}
\end{figure}

We illustrate the computation with an example shown in
Figure~\ref{fex}.
Suppose that edge lists contain edges in alphabetical order.
The first node is~1.
The first three iterations follow the unmatched edges
$a,b,c$, so that $\vc$ and the arrays $\vn$ and $\ve$ have
the following contents:
\[
\vc=4,
\quad
\vn=[1~~2~~3~~4],
\quad
\ve=[a~~b~~c].
\]
Node 4 has an empty \outlist, so that the computation
continues at line~18 (all line numbers refer to the main
method \fosm\ in Figure~\ref{fmain} unless specified
otherwise).
The matched edge is $m=45$ with endpoint $W=5$, and
$\shrink(c,45)$ is called in line~25.
Afterwards,
\[
45.\partner=c,
\quad
\vc=3,
\quad
\vn=[1~~2~~3],
\quad
\ve=[a~~b]
\]
and the reduced graph is shown in Figure~\ref{fex1}.
The nodes 3 and 5 have been united into the equivalence
class $\{3,5\}$ which has representative $5$ because
the \unite\ operation in Figure~\ref{funionfind} chooses the
second representative $Y$ if the original representatives
have equal rank.
The outgoing edges from node $5$ are $d$ and $h$ in that
order, because the outlist of $3$ has been appended to
that of~$5$.

\begin{figure}[hbt] 
\strut\hfill
\epsfysize9.6em
\epsfbox{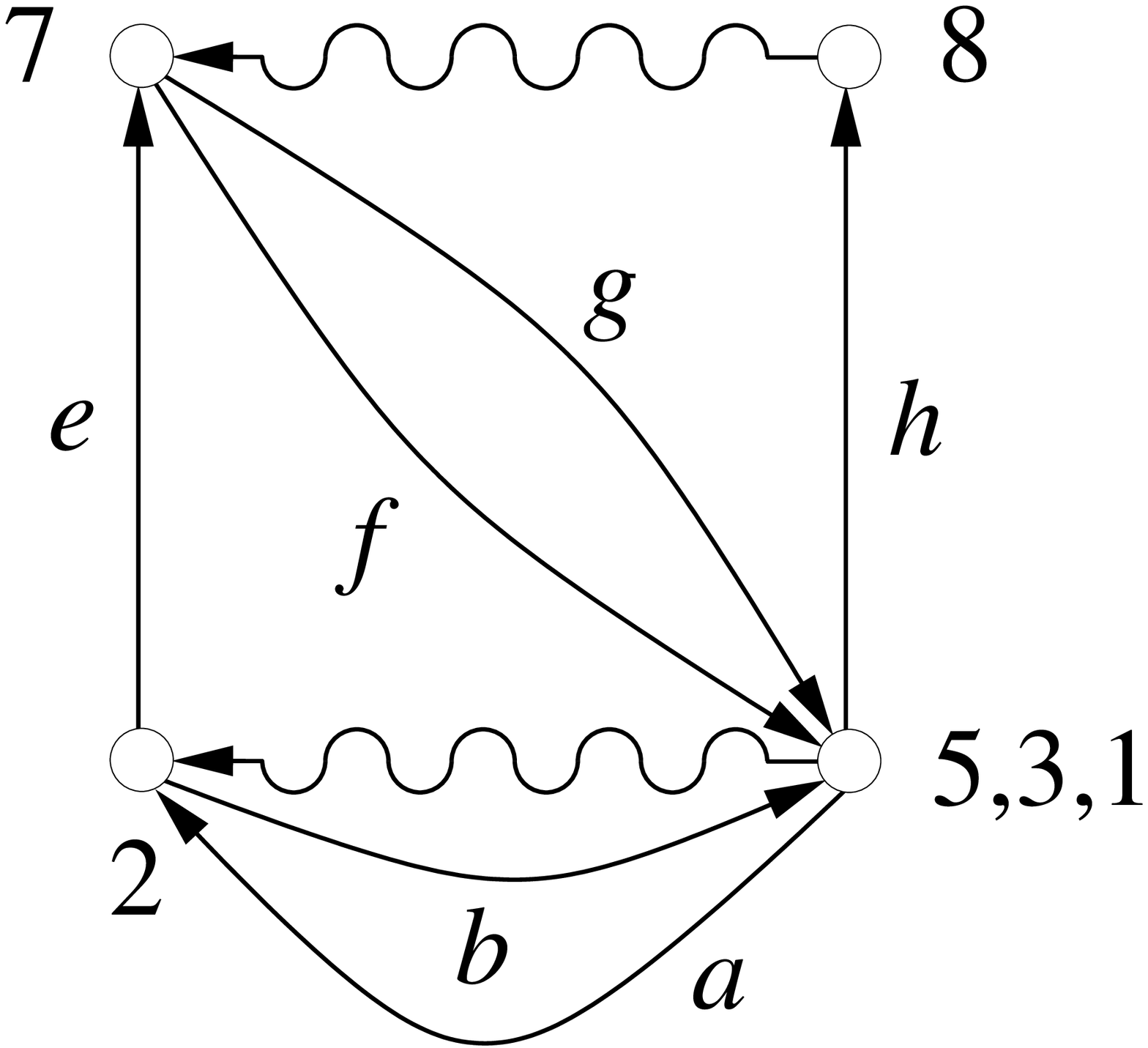} 
\hfill
\hfill
\epsfysize9.6em
\epsfbox{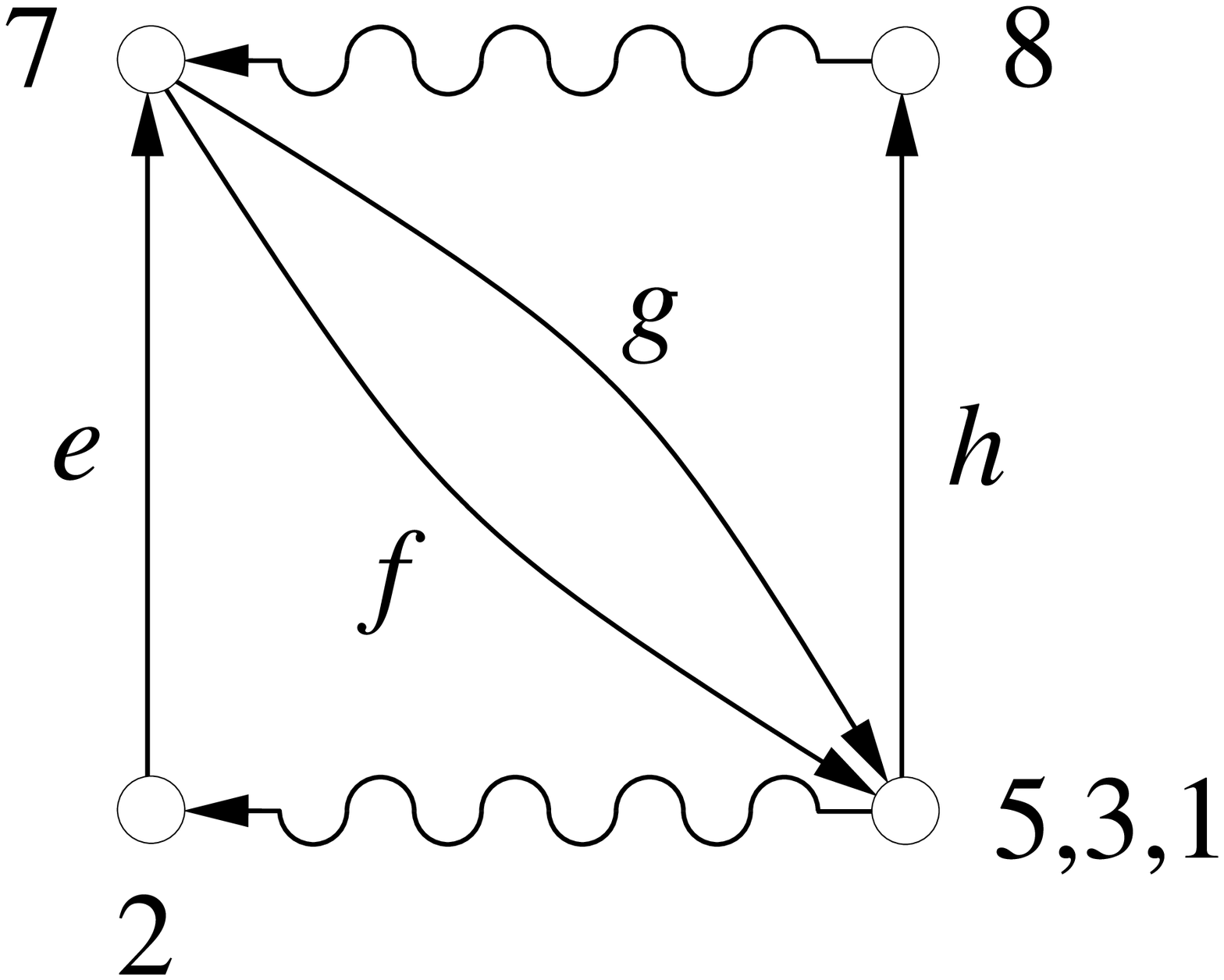} 
\hfill
\strut 
\caption{Left: after $\shrink(d,61)$, right: after 
$\checkvisited(1)$.
}
\label{fex2}
\end{figure}

In line 26, $\checkvisited(W)$ has no effect because
$W.\visited=0$.
In line 28, $V\SET \find(W)=W=5$,
so that after going back to step~{\bf B},
\[
\vc=3,
\quad
\vn=[1~~2~~5],
\quad
\ve=[a~~b].
\]
Line 11 then follows the unmatched edge $d$ with
\[
\vc=4,
\quad
\vn=[1~~2~~5~~6],
\quad
\ve=[a~~b~~d]
\]
after which again a contraction is needed, because
$V.\outlist$ is empty,
this time with $U,V,W=5,6,1$ and call to 
$\shrink(d,61)$.
In the \shrink\ method, $\unite(5,1)$ returns
$5,1$ because $5.\rank=1>0=1.\rank$.
The resulting graph, after line~25 is completed, is shown on the
left in Figure~\ref{fex2}, where  
\[
61.\partner=d,
\quad
W=1,
\quad 
\vc=3,
\quad
\vn=[1~~2~~5],
\quad
\ve=[a~~b].
\]
Now consider the call to $\checkvisited(W)$ in
line~26 and note that $W$ is still the old node $1$ used
before the contraction; recall that this is done because
that representative is possibly stored in the \vn\ array,
which it indeed is at index $W.\visited=1$.
The deletion of the detected cycle of unmatched edges in
lines 3 and~4 of \checkvisited\ (see Figure~\ref{fcheck})
then produces the reduced graph shown on the right in
Figure~\ref{fex2}.

\begin{figure}[hbt] 
\strut\hfill
\epsfysize4.6em
\epsfbox{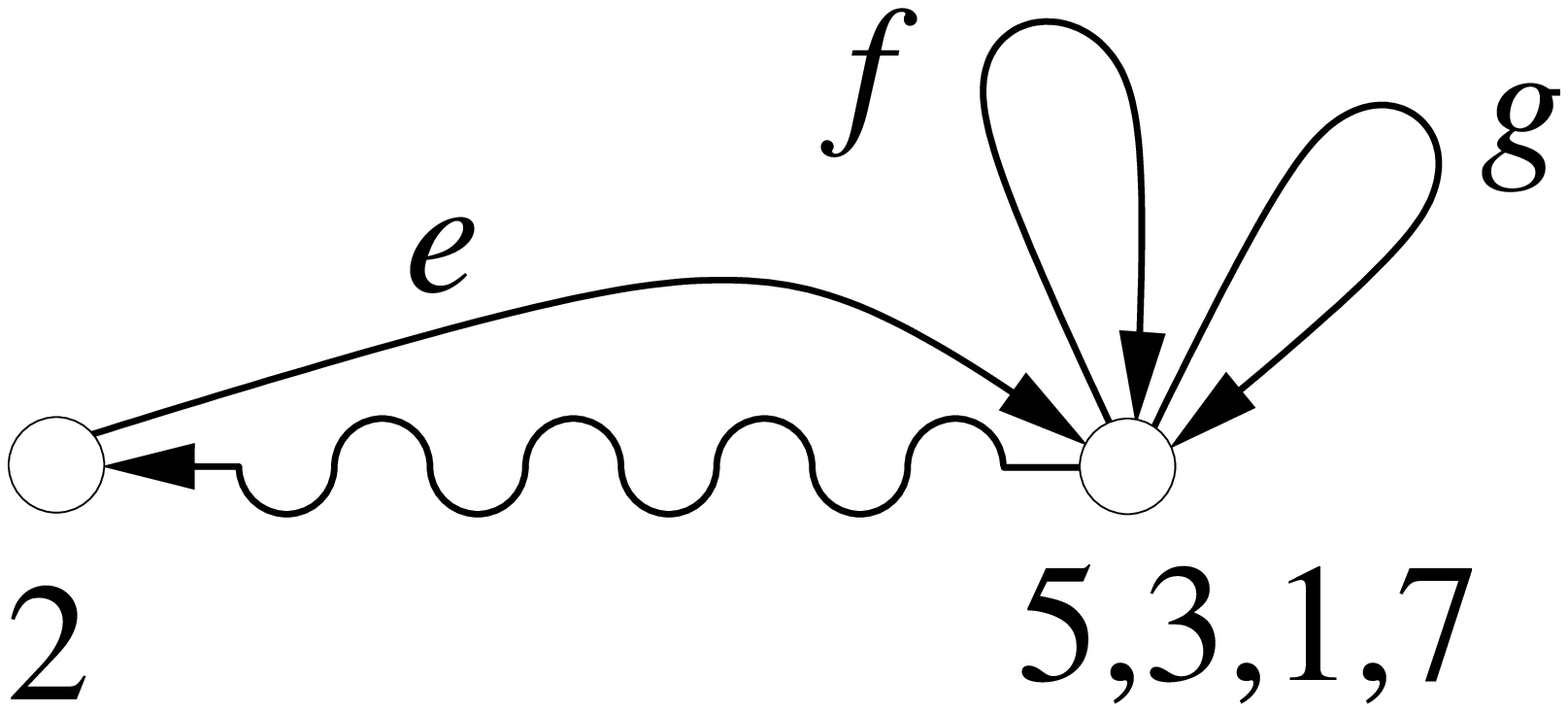} 
\hfill
\epsfysize4.6em
\epsfbox{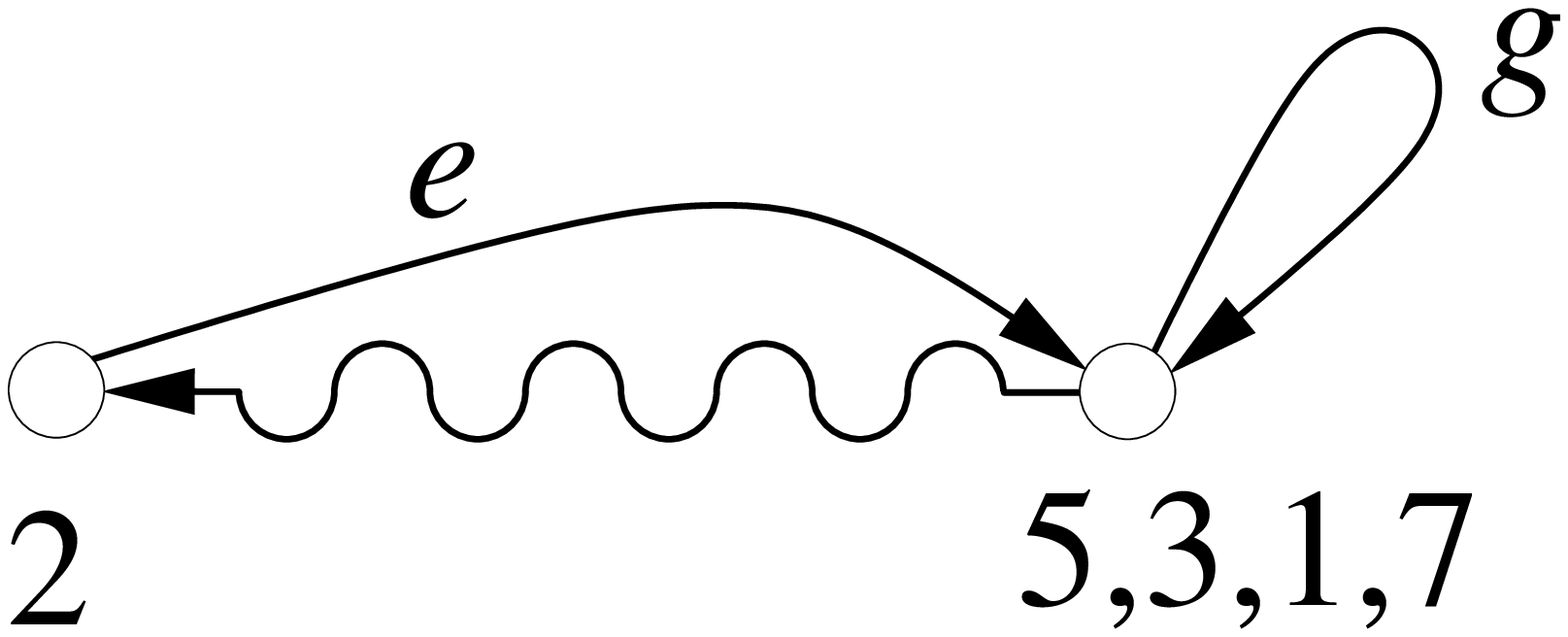} 
\hfill
\strut 
\caption{Left: reduced graph after the third contraction 
$\shrink(h,87)$, right: after $\checkvisited(7)$.
} 
\label{fex3}
\end{figure}

Normally, the next node $V$ would be $\find(W)$ in line~28.
However, the case $\vc=1$ applies (recall the reason that
the incoming matched edge of $\vn[1]$ has been removed by
the contraction), and so the computation continues via
line~31 to step~{\bf A}.
Suppose the first node is now~2.
Then the computation follows edges $e,f,h$ and reaches
node~8, with 
\[
\vc=4,
\quad
\vn=[2~~7~~5~~8],
\quad
\ve=[e~~f~~h].
\]
Contraction with $\shrink(h,87)$ gives the graph shown on
the left in Figure~\ref{fex3} where 
\[
87.\partner = h,
\quad
\vc=3,
\quad
\vn=[2~~7~~5],
\quad
\ve=[e~~f]
\]
and after $\checkvisited(7)$ the edge $f$ is removed,
resulting in 
\[
\vc=2,
\quad
\vn=[2~~7],
\quad
\ve=[e].
\]
This time, $\vc>1$, so with $V\SET\find(W)=5$ we go back via
line~29 to step~{\bf B}, after which
\[
\vc=2,
\quad
\vn=[2~~5],
\quad
\ve=[e]
\]
with the graph on the right in Figure~\ref{fex3}.

\begin{figure}[hbt] 
\strut\hfill
\epsfysize4.6em
\epsfbox{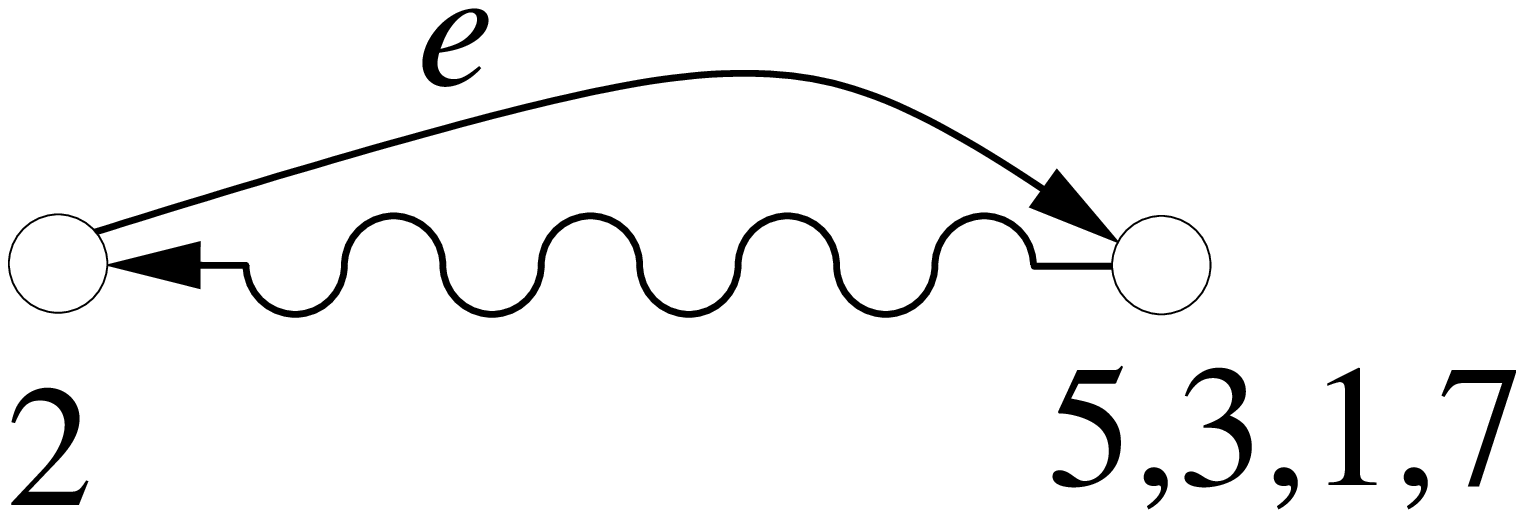} 
\hfill
\hfill
\epsfxsize9.0em
\epsfbox{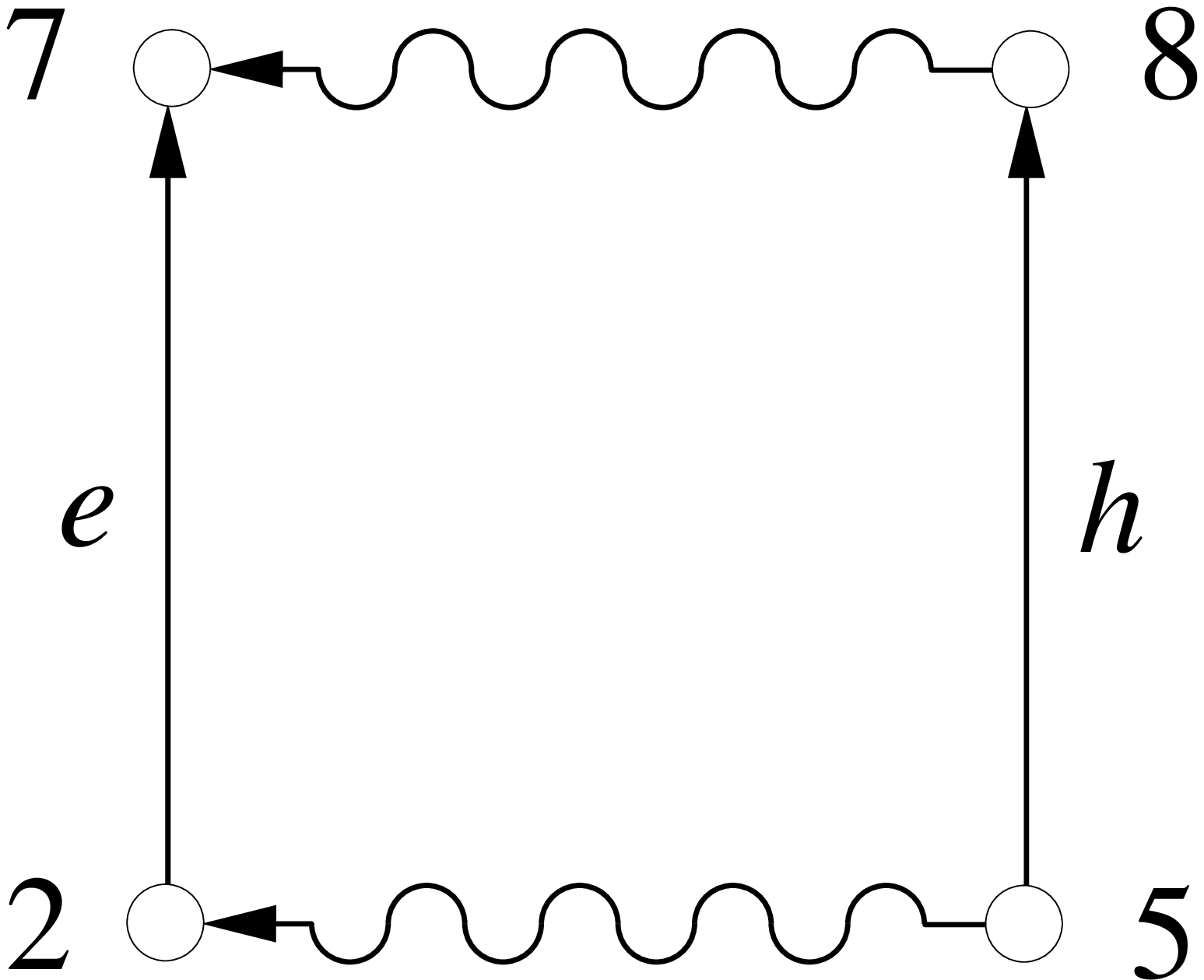} 
\hfill
\strut 
\caption{Left: graph after the removal of $g$, which has a
sign-switching cycle.
Right: cycle after the calls to \reconnect\ in \expandcycle.
}
\label{fex4}
\end{figure}

Now $V$ has only the outgoing unmatched edge~$g$, giving
in line 14 (where the last entry $\vn[vc]$ has not yet been
assigned)
\[
\vc=3,
\quad
\vn=[2~~5],
\quad
\ve=[e~~g]
\]
and removal of the edge $g$ gives the graph on the left of
Figure~\ref{fex4}.
At the next iteration, after line~9,
\[
\vc=2,
\quad
\vn=[2~~5],
\quad
\ve=[e]
\]
where $5.\outlist$ is empty, $m=52$, and now $W=2=U=\vn[1]$
in line~23.
Now the final stage of the algorithm is called in line~24
with $\expandcycle(e,m)$.  
The original endpoints of the two edges are (see
Figure~\ref{fex}):
$e.\tail=2$,
$e.\head=7$,
$m.\tail=3$,
$m.\head=2$.
Line~2 of Figure~\ref{freconnect} makes the call
$\reconnect(7,3,\{e,m\})$ which is nontrivial because
with $x,y=7,3$ we have $x\ne y$ in line~7 of
Figure~\ref{freconnect}. 
With $x.\origmatched=87$ and $87.\partner=h$,
we get $C=\{(e,52), (h, 87)\}$ (where $52$ is just our
current name for the matched edge, with its original
endpoints it is the edge $32$).
All other calls to $\reconnect(x,y,C)$ then have
no effect because $x=y$.
The resulting sign-switching cycle $C$ is shown on the right
in Figure~\ref{fex4}.  

In this example, every edge of the graph is visited during
the algorithm, and the reduced graph at the end consists
justs of the oppositely oriented unmatched and matched edge
that define a trivial sign-switching cycle.
The original graph in Figure~\ref{fex1} already has such an
edge pair in the form of $b,32$, which is not discovered by
the described run of the algorithm.
Here, not all matched edges and their partners that have
been removed by the \shrink\ operation are used (namely not
$45,c$ and $61,d$).
In other cases, the algorithm may also terminate with parts
of the graph left unvisited, or unmatched edges in the $\ve$
array that are not removed.

%\newpage 

\section*{References}
\frenchspacing
\parindent=-2em\advance\leftskip by2em
\parskip=.25ex plus.1ex minus.1ex
\small
\hskip\parindent
Balthasar, A. V. (2009),
Geometry and Equilibria in Bimatrix Games.
PhD Thesis, London School of Economics.  

Casetti, M. M., J. Merschen, and B. von Stengel (2010),
Finding Gale strings.
Electronic Notes in Discrete Mathematics 36, 1065--1072. 

Cayley, A. (1849),
Sur les d\'eterminants gauches.
Journal f\"ur die reine und angewandte Mathematik (Crel\-le's
Journal) 38, 93--96.  

Chen, X., and X. Deng (2006),
Settling the complexity of two-player Nash equilibrium.
Proc. 47th FOCS, 261--272.

Cormen, T. H., C. E. Leiserson, R. L. Rivest, and C. Stein (2001),
Introduction to Algorithms, Second Edition.
MIT Press, Cambridge, MA.

Cottle, R. W., and G. B. Dantzig (1970),
A generalization of the linear complementarity problem.
J. Combinatorial Theory 8, 79--90.  

Cottle, R. W., J.-S. Pang, and R. E. Stone (1992),
The Linear Complementarity Problem. Academic Press, San
Diego.

Daskalakis, C., P. W. Goldberg, and C. H. Papadimitriou (2009),
The complexity of computing a Nash equilibrium.
SIAM Journal on Computing 39, 195--259.  

Eaves, B. C., and H. Scarf (1976),
The solution of systems of piecewise linear equations.
Mathematics of Operations Research 1, 1--27.

Edmonds, J. (1965),
Paths, trees, and flowers.
Canadian Journal of Mathematics 17, 449--467.  

Edmonds, J. (2009),
Euler complexes. 
In: Research Trends in Combinatorial Optimization, eds. W.
Cook, L. Lovasz, and J. Vygen, Springer, Berlin, pp. 65--68.  

Edmonds, J., S. Gaubert, and V. Gurvich (2010),
Sperner oiks.
Electronic Notes in Discrete Mathematics 36, 1273--1280.

% Edmonds, J., and L. Sanit\`a (2010),
% On finding another room-partitioning of the vertices.
% Electronic Notes in Discrete Mathematics 36, 1257--1264.

Gale, D. (1963),
Neighborly and cyclic polytopes.
In: Convexity, Proc. Symposia in Pure Math., Vol. 7, ed. V.
Klee, American Math. Soc., Providence, Rhode Island,
225--232.

% Goldberg, P. W., C. H. Papadimitriou, and R. Savani (2011),
% The complexity of the homotopy method, equilibrium
% selection, and Lemke-Howson solutions.
% Proc. 52nd Annual IEEE Symposium on Foundations of Computer Science
% (FOCS), 67--76.

Grigni, M. (2001),
A Sperner lemma complete for PPA.
Information Processing Letters 77, 255--259.

Hilton, P. J., and S. Wylie (1967),
Homology Theory: An Introduction to Algebraic Topology.
Cambridge University Press, Cambridge.

Jacobi, C. G. J. (1827),
Ueber die Pfaffsche Methode, eine gew\"ohnliche line\"are
Differentialgleichung zwischen $2n$ Variabeln durch ein
System von $n$ Gleichungen zu integriren.
Journal f\"ur die reine und angewandte Mathematik (Crelle's
Journal) 2, 347--357.

Lax, P. D. (2007),
Linear Algebra and Its Applications.
Wiley, Hoboken, N.J. 

Lemke, C. E. (1965),
Bimatrix equilibrium points and mathematical
programming. Management Science 11, 681--689.

Lemke, C. E., and Grotzinger, S. J. (1976),
On generalizing Shapley's index theory to labelled
pseudomanifolds.
Mathematical Programming 10, 245--262.

Lemke, C. E., and J. T. Howson, Jr. (1964),
Equilibrium points of bimatrix games.
Journal of the Society for Industrial and
Applied Mathematics 12, 413--423.  

Lov\'asz, L., and M. D. Plummer (1986),
Matching Theory.
%Annals of Discrete Mathematics, Vol. 29.
North-Holland, Amsterdam.

McLennan, A., and R. Tourky (2010),
Simple complexity from imitation games.
Games and Economic Behavior 68, 683--688.  

Merschen, J. (2012),
Nash Equilibria, Gale Strings, and Perfect Matchings.
PhD Thesis, London School of Economics.
% \url{http://www.maths.lse.ac.uk/Personal/stengel/phds/#julian}

Morris, W. D., Jr. (1994),
Lemke paths on simple polytopes.
Mathematics of Operations Research 19, 780--789.

% Nash, J. (1951),
% Noncooperative games.  
% Annals of Mathematics 54, 286--295.

Papadimitriou, C. H. (1994),
On the complexity of the parity argument and other
inefficient proofs of existence.
Journal of Computer and System Sciences 48, 498--532.

Parameswaran, S. (1954),
Skew-symmetric determinants.
American Mathematical Monthly 61, 116.

Robertson, N., P. D. Seymour, and R. Thomas (1999),
Permanents, Pfaffian orientations, and even directed
circuits.
Annals of Mathematics 150, 929--975.  

Savani, R., and B. von Stengel (2006),
Hard-to-solve bimatrix games.
Econometrica 74, 397--429.  

% Scarf, H. (1967),
% The approximation of fixed points of a continuous mapping.
% SIAM Journal on Applied Mathematics 15, 1328--1343.

Shapley, L. S. (1974),
A note on the Lemke--Howson algorithm.
Mathematical Programming Study 1: Pivoting and Extensions,
175--189.  

Tarjan, R. E. (1975),
Efficiency of a good but not linear set union algorithm.
Journal of the ACM 22, 215--225.

Thomas, R. (2006),
A survey of Pfaffian orientations of graphs.
Proc. International Congress of Mathematicians, Madrid,
Spain, Vol. III, European Mathematical Society, Z\"urich,
963-–984.

Todd, M. J. (1972),
Abstract Complementary Pivot Theory.
PhD Dissertation, Yale University.

Todd, M. J. (1974),
A generalized complementary pivot algorithm.
Mathematical Programming 6, 243--263.

Todd, M. J. (1976),
Orientation in complementary pivot algorithms.
Mathematics of Operations Research 1, 54--66.  

Valiant, L. G. (1979),
The complexity of computing the permanent.
Theoretical Computer Science 8, 189--201.

Vazirani, V. V., and M. Yannakakis (1989),
Pfaffian orientations, 0-1 permanents, and even cycles in
directed graphs.
Discrete Applied Mathematics 25, 179--190.

% V\'egh, L. A., and B. von~Stengel (2012),
% Oriented Euler complexes and signed perfect matchings.
% arXiv: 1210.4694.

von~Stengel, B. (1999),
New maximal numbers of equilibria in bimatrix games.
Discrete and Computational Geometry 21, 557--568.

von~Stengel, B. (2002).
Computing equilibria for two-person games.
In: Handbook of Game Theory, Vol.~3, eds. R. J. Aumann and
S. Hart, North-Holland, Amsterdam, 1723--1759.

Ziegler, G. M. (1995),
Lectures on Polytopes. Springer, New York.

% \end{thebibliography} 
\end{document}